\documentclass[aps,prd,11pt,showkeys,nofootinbib]{revtex4-2}

\usepackage{subcaption}
\usepackage{amsmath,amssymb,amsfonts,amsthm}
\usepackage{graphicx}
\usepackage{hyperref}

\def\barr{\begin{array}}
\def\earr{\end{array}}

\def\ben{\begin{equation}}
\def\een{\end{equation}}
\def\bs{\begin{subequations}}
\def\es{\end{subequations}}
\def\bena{\begin{eqnarray}}
\def\eena{\end{eqnarray}}
\def\im{{\rm i}}

\begin{document} 
\title{\bf Inflation in unimodular loop quantum cosmology}

\author{Steffen Gielen\footnote{Corresponding author}, Rita B. Neves}
\affiliation{School of Mathematical and Physical Sciences, University of Sheffield, Hicks Building, Hounsfield Road, Sheffield S3 7RH, United Kingdom}
\email{s.c.gielen@sheffield.ac.uk}
\email{rita.neves@sheffield.ac.uk}

\begin{abstract}
We study inflation in the setting of unimodular loop quantum cosmology, where time evolution is defined in unimodular time rather than with respect to a free, massless scalar field as is standard in loop quantum cosmology. The unimodular setting leads to a natural Schr\"odinger time evolution in a time coordinate with clear geometric meaning, defined independently of any particular matter content; an inflaton can be included but is not needed as a clock. We review the unimodular version of loop quantum cosmology and comment on possible connections to full (unimodular) loop quantum gravity. Then, focusing on semiclassical effective equations, we derive analytical solutions in simple cases such as a constant potential, emphasising the use of a unimodular time coordinate. We also discuss numerical solutions for phenomenologically interesting cases such as a quadratic potential and Starobinsky inflation, comparing different possible choices of initial conditions. In particular, we show that choosing an $\alpha$-attractor potential allows for models of a bounce either dominated by kinetic or potential energy, which are compatible with observations while potentially including observable imprints of the quantum-gravity regime. 
\end{abstract}


\maketitle

\section{Introduction}

In modern cosmology it is widely assumed that the very early Universe went through a period of rapid, nearly exponential expansion known as inflation, the simplest mechanism for which is the presence of a scalar field with a suitable, nearly flat potential \cite{Baumann_2022}. While inflationary models can successfully account for a wide range of observed phenomena\footnote{See, e.g., \cite{Martin:2013tda} for a heroic attempt to catalogue and test hundreds of possible models of inflation, following the initial release of PLANCK data in 2013.}, including a prediction of nearly scale-invariant, adiabatic, Gaussian scalar perturbations, a theory based only on general relativity and inflation is necessarily incomplete as it cannot account for the initial state of the Universe at the Big Bang, which is singular \cite{Borde:2001nh}. A natural suggestion for extending our incomplete picture of cosmology is that such an extension can be found within a quantum theory of gravity. A range of approaches to quantum gravity indeed replace the initial singularity by a non-singular description, often (but not necessarily) in terms of a Big Bounce \cite{Ashtekar:2011ni,Oriti:2016qtz,Brahma:2021tkh}. In such a nonsingular scenario, one may hope for the existence of an alternative mechanism for generating the observed cosmological inhomogeneities, so that no additional inflaton field is needed \cite{Khoury:2001wf,Brandenberger:2016vhg,Marchetti:2025jze}, or one may consider the need to resolve the Big Bang singularity and the need to explain the observed large-scale structure of the Universe as disconnected problems, so that a desired inflationary model can simply be added to an existing quantum theory of gravity. The latter approach is the one adopted in loop quantum cosmology (see, e.g., \cite{Bojowald:2011iq}) and will also be used in this paper. While perhaps less ambitious in its scope for providing simple, unified answers for how the Universe originated, such an approach provides more flexibility in studying different models and hence potentially accounting for observational data up to the very high precision that is required today. 

Loop quantum cosmology \cite{Ashtekar:2011ni} is constructed by applying quantisation methods of loop quantum gravity \cite{Ashtekar:2004eh} to the context of homogeneous, isotropic cosmology, i.e., to a minisuperspace model. Crucially, features of loop quantum gravity such as the discreteness of geometric observables (areas, volumes, etc) mean that the resulting quantisation is inequivalent to the Wheeler--DeWitt quantisation used traditionally in quantum cosmology \cite{Misner:1969ae,GIELEN2025520}, and leads to different physical predictions, most notably the universal appearance of a Big Bounce. As a canonical quantisation of gravity (even if only in a symmetry-reduced setting), loop quantum cosmology models face the problem of time \cite{Isham:1992ms}, in that the quantum theory does not have a preferred time parameter with respect to which evolution can be defined. This problem is addressed by working within a Dirac (constraint) quantisation in which one defines a physical from a kinematical Hilbert space by projecting onto the states solving the Hamiltonian constraint, in a process known as group averaging (again, see \cite{Ashtekar:2011ni}; a simpler example of group averaging can be found in \cite{Hohn:2018iwn}). For such a construction to be possible explicitly, it is generally necessary to have a degree of freedom that can serve as a ``clock'' and effectively define the notion of time evolution. In loop quantum cosmology this variable is taken to be a free, massless scalar field, which indeed evolves monotonically and can hence label different instants of time uniquely. Such an approach is relatively standard in quantum cosmology and allows formulating physical questions in ``relational'' terms, by characterising the evolution of the Universe with respect to a physical degree of freedom.

This standard approach however raises several questions. A fundamental one is whether the resulting theory is independent of this choice of clock variable, i.e., whether a different theory constructed with the same methods, but based on a different choice of clock, would be equivalent to standard loop quantum cosmology. While there is so far little work on this question, in the case of Wheeler--DeWitt quantum cosmology models constructed with different clocks (even within the framework of group averaging) are inequivalent, and can make radically different predictions \cite{Gielen:2020abd}. In particular, using a so-called ``slow'' clock associated to a perfect fluid (e.g., dust or dark energy) allows achieving singularity resolution in a wide range of contexts without the need for a loop quantisation \cite{Gryb:2018whn,Gielen:2022tzi,Gielen:2024lpm,Gielen:2025ovv}. From a more phenomenological point of view, it might be interesting to compare the results of loop quantum cosmology models for cosmological perturbations to other models constructed using different clocks (see \cite{Giesel:2020raf} for some work in this direction).

A second objection that a cosmologist might have is whether adding free, massless scalar fields is realistic when there is no evidence of the existence of such fields in our Universe. This criticism would apply to many models of quantum cosmology, for example models based on the popular Brown--Kucha\v{r} dust \cite{Brown:1994py}, in which matter fields are designed to be good (rods and) clocks rather than realistic descriptions of our Universe. One might think of a scalar field as representing the inflaton, which however requires a rather specific potential. In phenomenological studies of inflation in loop quantum cosmology \cite{Agullo:2013ai}, the usual idea is that the bounce is dominated by kinetic rather than potential energy in the scalar field, so that in the Planck-scale regime it would be justified to approximate the inflaton as a free scalar field. The period of kinetic domination is expected to end quickly, but if this is far away from the Planck regime one might then be content with studying the impact of the inflationary potential within standard cosmology rather than an embedding into a theory derived from quantum gravity, again viewing singularity resolution and inflation as separate problems. 

In this paper we address both possible questions by studying a unimodular version of loop quantum cosmology, first proposed in \cite{Chiou:2010ne} (and also studied in \cite{Gielen:2024qml}). Unimodular versions of general relativity \cite{Buchmuller:1988yn,Finkelstein:2000pg,Henneaux:1989zc} have long been discussed as a possible way of addressing the problem of time \cite{Unruh:1989db,Kuchar:1991xd}. While locally equivalent to standard general relativity, such theories include an additional {\em global} degree of freedom, associated to the cosmological constant which appears as an integration constant rather than a fundamental constant of Nature. This degree of freedom is very similar to a total energy, and naturally conjugate to a time parameter, the time fixed by the unimodular condition. This time is proportional to the total four-volume (of the entire Universe or of a finite spatial region of interest), and hence has a natural geometric interpretation. In a minisuperspace context, the additional degree of freedom in unimodular gravity is indistinguishable from a perfect fluid with equation of state $p=-\rho$, representing dark energy (see, e.g., \cite{Gielen:2022tzi}). Hence, in such symmetry-reduced models, it provides an obvious candidate for a clock, which transforms the Wheeler--DeWitt equation into a Schr\"odinger equation in unimodular time, as already anticipated in \cite{Unruh:1989db} (see also \cite{Gryb:2018whn} for similar conclusions from a different conceptual angle). One can then add a scalar field with arbitrary potential, for instance to model inflation, but the scalar would no longer be used as a clock. This is the perspective we want to advocate in this paper. We should stress that the time coordinate provided by unimodular formulations of gravity is tied to a particular foliation, i.e., it labels spatial hypersurfaces uniquely {\em once a foliation has been fixed}. This is the essence of the criticism of Kucha\v{r} \cite{Kuchar:1991xd}. Hence, working within unimodular gravity does not provide a universal resolution of the problem of time outside of simple settings such as homogeneous cosmology (or static and spherically symmetric black holes \cite{Gielen:2025ovv}), where a preferred foliation exists by assuming a large amount of symmetry. For the purposes of this paper, we content ourselves with the setting of homogeneous cosmology, where we want to study unimodular loop quantum cosmology as an alternative to the conventional models. Our study is largely exploratory and emphasises technical, conceptual and possible phenomenological aspects, hoping to inspire future work on unimodular versions of loop quantum cosmology. In the approach to phenomenology, we emphasise the possibility of a bounce dominated by potential energy in the scalar field, since the approximation by a free scalar field (with only kinetic energy) is no longer needed in our approach.

In Section \ref{LQCsec} we introduce the basic ideas behind unimodular formulations of loop quantum gravity and loop quantum cosmology, emphasising possible connections between a unimodular minisuperspace model and a possible full canonical quantisation (assuming it could be defined). We focus on a homogeneous, isotropic and spatially flat Friedmann--Lema\^{i}tre--Robertson--Walker (FLRW) Universe in which the matter content is given by a scalar field with arbitrary potential. The cosmological constant arises as an integration constant, akin to the total energy of a mechanical system, and is conjugate to a preferred time parameter $T$. While it would be interesting to study the quantum theory in full detail, here we focus on effective semiclassical equations which can be obtained from evaluating loop-corrected quantum operators in semiclassical states, or simply by importing loop (holonomy) corrections into the classical dynamics. Section \ref{classsol} shows the classical solutions of such a model (i.e., of the model without holonomy corrections), mostly in simple cases such as a constant potential, although it is possible to derive at least special solutions in more general settings. Some of these solutions can be found elsewhere in the literature (e.g., \cite{Gielen:2020abd}) but are stated for completeness, also in terms of both unimodular time and the scalar field time used traditionally in loop quantum cosmology. Section \ref{effectivesol} shows exact solutions in unimodular effective loop quantum cosmology for the case of a constant scalar field potential. Here in the general case exact solutions can only be derived in cosmic time, and quantities such as the volume evolving in unimodular time can be discussed in terms of parametric plots. Many of these solutions have not been derived in as much detail in previous literature. The volume can be given as a function of scalar time in terms of Jacobi elliptic functions. Finally, in Section \ref{numericalsol} we turn to the more interesting cases of a nontrivial scalar field potential, focusing on a quadratic potential, Starobinsky inflation and $\alpha$-attractors. Here, the use of unimodular time means that it is not necessary for the scalar field to evolve monotonically everywhere, since it is no longer used as a clock. We show a range of numerical solutions with different initial conditions, corresponding to a bounce that can be kinetically dominated or dominated by potential energy. While similar solutions have been given in the literature, we show some cases that are usually not considered and we emphasise, again for the first time, the use of unimodular time in the study of inflationary models in loop quantum cosmology. Finally, we give some conclusions in Section \ref{conclsec}. 

\section{Unimodular loop quantum cosmology}
\label{LQCsec}

In the standard formulation of canonical loop quantum gravity, the dynamics of general relativity are expressed in terms of conjugate variables $(A_a^i, \tilde{E}_j^b)$ representing the Ashtekar--Barbero connection and its conjugate, the densitised triad. In those variables and in canonical form, the Einstein--Hilbert action can be replaced by the action \cite{Ashtekar:2004eh}
\begin{align}
  S_{{\rm LQG}}[A,\tilde{E},\alpha,V,N]&=\frac{1}{\kappa\gamma}\int {\rm d}^4 x\bigl[\dot{A}^i_a \tilde{E}_i^a - \alpha^i\,D_a\tilde{E}^a_i
 -V^a\,F^i_{ab}\tilde{E}^b_i
 \\& \qquad-N\,\frac{\gamma\tilde{E}^a_i\tilde{E}^b_j}{2\sqrt{\det \tilde{E}}}\left({\epsilon^{ij}}_k F^k_{ab}-2(1+\gamma^2)K^i_{[a}K^j_{b]}\right)\bigr]\,. \nonumber
\end{align}
Here $D_a$ is a covariant derivative using $A_a^i$, $F^i_{ab}$ is the curvature 2-form of $A_a^i$, $K^i_a$ is the extrinsic curvature associated to the fundamental variables (defined via $A^i_a = \Gamma^i_a + \gamma K^i_a$, where $\Gamma^i_a$ is the Levi-Civita connection associated to the triad $\tilde{E}_j^b$) and $\gamma$ is the Barbero--Immirzi parameter. We also set $\kappa=8\pi G$.

The variables $\alpha^i$, $V^a$ and $N$ appear as Lagrange multipliers imposing the Gauss, diffeomorphism (momentum) and Hamiltonian constraints, which have to be implemented in the quantum theory as constraints on $(A_a^i, \tilde{E}_j^b)$. While the Gauss and diffeomorphism constraints can be implemented kinematically at the level of spin-network states, the Hamiltonian constraint is crucial for implementing the dynamics of general relativity and for defining the physical Hilbert space of loop quantum gravity \cite{Ashtekar:2004eh}. Various proposals exist, starting from the seminal work of \cite{Thiemann_QSD} up to more recent developments \cite{Assanioussi:2015gka,Yang:2015zda,Varadarajan:2022dgg}. Identifying a space of (physical) states annihilated by a proposed constraint, with its physical inner product, is an extremely difficult problem.

In this standard formulation the Hamiltonian constraint arises from varying with respect to the lapse $N$. In a unimodular approach, the lapse is fixed by the unimodular condition on the spacetime volume element, and so there is no Hamiltonian constraint but a true Hamiltonian. This would initially seem to bypass the need to construct the kernel of a constraint and find a new physical inner product, as it does in symmetry-reduced cosmological models. If we assume that the volume form is fixed, i.e.,
\begin{equation}
N\,\sqrt{\det\tilde{E}}=\tilde{N}_0
\end{equation}
where $\tilde{N}_0$ is a given background field, we obtain the unimodular loop quantum gravity action
\begin{align}
  S_{{\rm UniLQG}}[A,\tilde{E},\alpha,V;\tilde{N}_0]&=\frac{1}{\kappa\gamma}\int {\rm d}^4 x\bigl[\dot{A}^i_a \tilde{E}_i^a - \alpha^i\,D_a\tilde{E}^a_i
 -V^a\,F^i_{ab}\tilde{E}^b_i
 \\& \qquad-\tilde{N}_0\,\frac{\gamma\tilde{E}^a_i\tilde{E}^b_j}{2\det \tilde{E}}\left({\epsilon^{ij}}_k F^k_{ab}-2(1+\gamma^2)K^i_{[a}K^j_{b]}\right)\bigr]\,. \nonumber
\end{align}
This action is a generalisation of unimodular actions that can be derived from unimodular formulations of Pleba\'nski gravity \cite{Smolin:2010iq,Gielen:2024gfb}. Specifically, the {\em canonical preferred-volume action} derived in \cite{Gielen:2024gfb} takes the form
\begin{equation}
S_{{\rm CanPV}}[A,\tilde{E},\alpha,V;\tilde{N}_0] =\frac{1}{{\rm i}\kappa}\int {\rm d}^4 x\bigl[\dot{A}^i_a\tilde{E}_i^a + \alpha^i\,D_a\tilde{E}^a_i -V^a\left(F^i_{ab}\tilde{E}^b_i-A^i_a D_b\tilde{E}^b_i\right)-\tilde{N}_0\,\frac{{\epsilon^{ij}}_k F^k_{ab}\tilde{E}^a_i\tilde{E}^b_j}{2\det\tilde{E}}\bigr]\,.
\end{equation}
This corresponds to the special case where $\gamma=\pm{\rm i}$ and we work with self-dual complex variables. These variables would require reality conditions in the quantum theory and are hence usually avoided, even though their spacetime geometric interpretation is clearer than that of the Ashtekar--Barbero connection for real $\gamma$.

We see that now indeed there is initially no Hamiltonian constraint, since we no longer have the Lagrange multiplier $N$. However, the ``true Hamiltonian'' part of the total Hamiltonian has a nonvanishing Poisson bracket with the diffeomorphism constraint, leading to a secondary constraint
\begin{equation}
\mathcal{K}_a = \partial_a \left( \frac{\tilde{E}^b_i\tilde{E}^c_j}{2\kappa\det \tilde{E}}\left({\epsilon^{ij}}_k F^k_{bc}-2(1+\gamma^2)K^i_{[b}K^j_{c]}\right) \right) \approx 0\,.
\label{eq:newconst}
\end{equation}
This implies
\begin{equation}
\frac{\tilde{E}^a_i\tilde{E}^b_j}{2\kappa\det \tilde{E}}\left({\epsilon^{ij}}_k F^k_{ab}-2(1+\gamma^2)K^i_{[a}K^j_{b]}\right) +\frac{\Lambda}{\kappa} \approx 0
\label{eq:lambdaconst}
\end{equation}
where $\Lambda$ is some unspecified integration constant -- initially only constant in space, but then also constant in time from self-consistency of the Hamiltonian evolution. $\Lambda$ is a single additional global degree of freedom in unimodular gravity compared to general relativity.

The secondary constraints $\mathcal{K}_a$ can be added to the theory by adding a term $\tilde{T}^a\mathcal{K}_a$ to the Hamiltonian, where $\tilde{T}^a$ is a Lagrange multiplier vector density, as done in \cite{Gielen:2024gfb}.

As usual, the structure of the constraints simplifies dramatically in symmetry-reduced models, particularly in the very simplest case of flat FLRW cosmology. Unimodular loop quantum cosmology is based on the same variables as in the standard approach, with the main difference that the Hamiltonian constraint is replaced by the symmetry-reduced version of (\ref{eq:lambdaconst}). Concretely, one assumes that spacetime is of the form $\Sigma\times\mathbb{R}$, where the spatial hypersurfaces $\Sigma$ are often taken to be non-compact $\mathbb{R}^3$ so that integrations have to be restricted to a ``fiducial cell'' of coordinate volume $\mathcal{V}$. For flat FLRW cosmology, the Ashtekar--Barbero variables take the form
\begin{equation}
    A^i_a = \mathcal{V}^{-1/3}\,c(t)\,{}^0\omega^i_a\,,\quad \tilde{E}^a_i = \mathcal{V}^{-2/3}\sqrt{{}^0 q}\,p(t)\,{}^0 e^a_i
    \label{AshtekarAnsatz}
\end{equation}
where ${}^0\omega^i_a$ is a fiducial co-triad such that
\begin{equation}
    {}^0 q_{ab}=({}^0\omega^i_a)({}^0\omega^j_b)\delta_{ij}
\end{equation}
is a fiducial flat background metric. The physical spatial metric is then given by
\begin{equation}
    q_{ab}=a^2(t)\, {}^0 q_{ab} = {}^0 q_{ab}\,|p(t)|\mathcal{V}^{-2/3}\,.
    \label{eq:physmetric}
\end{equation}
These definitions imply that the physical volume of the fiducial cell is proportional to $|p(t)|^{3/2}$ and does not depend on the coordinate volume $\mathcal{V}$; both $c$ and $p$ are invariant under a redefinition of the fiducial cell. These variables satisfy
\begin{equation}
    \{c,p\} = \frac{\kappa\gamma }{3}\,.
\end{equation}

They are the standard kinematical variables of loop quantum cosmology \cite{Banerjee:2011qu}. Substituting (\ref{AshtekarAnsatz}) into (\ref{eq:lambdaconst}) and using the definition $A^i_a=\Gamma^i_a+\gamma K^i_a$ of the Ashtekar--Barbero connection, together with the fact that here the spatial Levi-Civita connection $\Gamma^i_a$ vanishes, yields the simple form
\begin{equation}
   -\frac{3}{\kappa\gamma^2}\frac{c^2}{|p|} + \frac{\Lambda}{\kappa} \approx 0 \,.
\end{equation}
$\Lambda$ is a dynamical degree of freedom conjugate to a clock $T$ which one may define to satisfy
\begin{equation}
    \{T,\Lambda\} = \kappa\,.
\end{equation}
This clock variable can be obtained more systematically from symmetry reduction of the Henneaux--Teitelboim version of unimodular gravity \cite{Henneaux:1989zc} to flat FLRW cosmology, as shown in \cite{Chiou:2010ne}.

Finally, given that we are interested in cosmological applications and inflation in particular, we may add a scalar field as matter. Its Lagrangian with unimodular volume form is
\begin{equation}
  \mathcal{L}_\phi = \tilde{N}_0\left(\frac{\det\tilde{E}}{2\tilde{N}_0^2}\dot\phi^2-V(\phi)\right)
\end{equation}
and the constraint now becomes
\begin{equation}
   -\frac{3}{\kappa\gamma^2}\frac{c^2}{|p|} + \frac{\Lambda}{\kappa} +\frac{\pi_\phi^2}{2|p|^3}+V(\phi)\approx 0
   \label{eq:constraint}
\end{equation}
in agreement with \cite{Chiou:2010ne}. Here $\pi_\phi$ is the conjugate momentum to the scalar field $\phi$, with $\{\phi,\pi_\phi\}=1$. In the symmetry-reduced setting, this constraint contains the entire dynamics of the theory. 

In the ``obvious'' gauge fixing of the theory where $T$ is used as a time, the constraint can again be replaced by the true Hamiltonian
\begin{equation}
\mathcal{H}_0 =  -\frac{3}{\kappa\gamma^2}\frac{c^2}{|p|}  +\frac{\pi_\phi^2}{2|p|^3}+V(\phi)
\end{equation}
which generates evolution in $T$; quantisation leads to a standard Schr\"odinger equation in $T$.

\subsection{Quantum theory}

The phase-space variables defined in the classical theory should now be represented as operators on a kinematical Hilbert space, and physical states should be expected to satisfy a quantum analogue of (\ref{eq:constraint}). This could be done in the standard way, e.g., by defining a representation in which $p$, $\Lambda$ and $\pi_\phi$ act as derivatives on a space of wavefunctions $\Psi(c,T,\phi)$. In loop quantum cosmology one uses the additional assumption that the connection $c$ is not represented as an operator; instead, only its holonomies are well-defined. This mimics what happens in loop quantum gravity, where likewise only finite holonomies of the connection are defined as operators.

Concretely, define operators $\hat{\mathcal{N}}_\mu$ by \cite{Banerjee:2011qu}
\begin{equation}
   \hat{\mathcal{N}}_\mu = \widehat{e^{\frac{{\rm i}}{2}\mu c}}\,, \quad [\hat{\mathcal{N}}_\mu,\hat{p}]=-\frac{\hbar\kappa\gamma}{6}\,\mu\,\hat{\mathcal{N}}_\mu
\end{equation}
with the commutator following from $[\hat{\mathcal{N}}_\mu,\hat{p}]={\rm i}\hbar\widehat{\{e^{\frac{{\rm i}}{2}\mu c},p\}}$. In a representation in which $\hat{p}$ is diagonal, $\hat{\mathcal{N}}_\mu$ generates shifts in the eigenvalues:
\begin{equation}
   \hat{p}|p\rangle=p|p\rangle\qquad \Rightarrow\quad\hat{p}\left(\hat{\mathcal{N}}_\mu|p\rangle\right) = \left(p+\frac{\hbar\kappa\gamma}{6}\mu\right)\left(\hat{\mathcal{N}}_\mu|p\rangle\right)\,.
\end{equation}
In this construction, $\mu$ can be a constant or a function of $p$, fixed by physical arguments and heuristic connection with loop quantum gravity (see below).

The matter variables $\phi$ and $\pi_\phi$ are usually {\em not} quantised as they would be in loop quantum gravity (where they should be represented as ``point holonomies'' \cite{Thiemann:1997rq}), but instead one assumes a standard representation with $[\hat\phi,\hat{\pi}_\phi]={\im}\hbar$. We will assume the same for the ``dark energy'' variables with $[\hat{T},\hat{\Lambda}]={\im}\hbar\kappa$, as was also done in \cite{Chiou:2010ne}.

\subsection{Remarks on connection to full loop quantum gravity}

In a canonical quantisation programme for general relativity, adopting a unimodular approach means that the difficult problem of imposing the Hamiltonian constraint is replaced by the seemingly equally difficult problem of imposing its spatial derivative (\ref{eq:newconst}). In spatially homogeneous models, (\ref{eq:newconst}) is trivial since spatial derivatives are forbidden by symmetry. Hence, when studying isotropic or anisotropic cosmology or the homogeneous interior of a Schwarzschild black hole, unimodular extensions of loop quantum gravity provide technical advantages over the usual constrained formulation. We have one constraint less than in standard general relativity, but in such homogeneous situations there was only a single constraint to begin with. On the other hand, such homogeneous models are also the cases where the Hamiltonian constraint {\em can} already be implemented in the standard approach, at least after addition of suitable reference matter (such as dust or the usual free massless scalar field).

While beyond the scope of our work here, it might be worthwhile to construct implementations of (\ref{eq:newconst}) in canonical loop quantum gravity, since this has not been done yet. The usual constructions of a scalar constraint operator (e.g., \cite{Assanioussi:2015gka}) take the form
\begin{equation}
    \hat{C}(N) = \sum_{v\in \Gamma} N(v)\hat{C}_v
\end{equation}
where $\Gamma$ is the graph defining a loop quantum gravity spin-network state, $v$ is a vertex or node of the graph, and $\hat{C}_v$ is a suitably defined operator acting locally at each vertex $v$. Such a definition represents the discrete version of the classical smeared constraint 
\begin{equation}
C[N]=\int {\rm d}^3 x\;N(x)C(x)
\end{equation}
where $C(x)$ is the Hamiltonian constraint and $N(x)$ is a smearing function which can be seen as representing the lapse. Classically, we need $C[N]\approx 0$ for all choices of $N$, so quantum-mechanically we should require physical states to be annihilated by $\hat{C}(N)$ for arbitrary choices of $N(v)$.

In the unimodular setting, we do not need to impose a Hamiltonian constraint but rather its spatial derivative (\ref{eq:newconst}), or alternatively (\ref{eq:lambdaconst}) where $\Lambda$ is a global degree of freedom. One idea could be to assign $\Lambda$ as an additional label to a spin network and define a new scalar constraint operator
\begin{equation}
    \hat{C}(N,\Lambda) = \sum_{v\in \Gamma} N(v)\hat{C}_{v,\Lambda}
\end{equation}
where $\hat{C}_{v,\Lambda}$ is a proposal for a loop quantum gravity scalar constraint operator with cosmological constant $\Lambda$. A suitable definition of such an operator would correspond to imposing the constraint
\begin{equation}
\frac{\tilde{E}^a_i\tilde{E}^b_j}{\sqrt{\det \tilde{E}}}\left({\epsilon^{ij}}_k F^k_{ab}-2(1+\gamma^2)K^i_{[a}K^j_{b]}\right) \approx \Lambda\,\sqrt{\det\tilde{E}}
\end{equation}
which has the scalar density weight $+1$ needed in loop quantum gravity. The right-hand side should be discretised using the volume operator, as one would do for the standard Hamiltonian constraint with cosmological constant $\Lambda$. We then still have the major issue of defining a physical inner product for states satisfying the new constraints. If this could be done, the expectation would be that the total Hilbert space of the theory is a direct sum of Hilbert spaces $\mathcal{H}_\Lambda^{{\rm LQG}}$ each associated with a given value of $\Lambda$. Future work would be needed to establish whether this tentative proposal can be turned into a consistent quantum theory, or whether it alleviates any of the technical issues in canonical loop quantum gravity.

The perspective we will take here is that the main advantage of unimodular formulations of general relativity is to provide an additional clock degree of freedom with respect to which unitarity can be defined. This is in particular true for symmetry-reduced models where the usual constraint is replaced by an equation generating evolution in unimodular time. No additional reference clock matter, such as a free massless scalar field, is needed. Scalar fields, such as the inflaton, can instead be used as physical degrees of freedom relative to the unimodular clock. 

\subsection{Improved dynamics and new canonical variables}

As in most of the literature, we will use the improved dynamics prescription \cite{improve} in which the function $\mu$ appearing in the definition of the operators $\hat{\mathcal{N}}_\mu$ is taken to be
\begin{equation}
\bar\mu = \sqrt{\frac{\Delta}{|p|}}\,, \qquad \Delta = \frac{\sqrt{3}}{2}\hbar\kappa\gamma\,,
\label{eq:improdyn}
\end{equation}
where $\Delta$ is a minimal Planck-scale area corresponding to the ``area gap'', the lowest non-zero eigenvalue for areas in loop quantum gravity. The somewhat heuristic idea behind this prescription is that the physical length associated to a minimal loop, whose holonomy is used when replacing the continuum connection, should remain fixed. Hence, the coordinate length associated to it should scale as $1/\sqrt{|p|}$ (or as the inverse of the cosmological scale factor) in an evolving FLRW geometry. $\Delta$ is a new scale in the theory, motivated by loop quantum gravity calculations, and in the following we could replace $\Delta$ by its explicit form given in (\ref{eq:improdyn}). However, one might also see $\Delta$ as a free parameter, determined perhaps by observation \cite{Ashtekar:2021kfp}, so we will leave $\Delta$ as an independent constant in the expressions below.

The operator $\hat{\mathcal{N}}_{\bar\mu}$ then depends on the combination $c/\sqrt{|p|}$, and it is convenient to introduce new canonical variables
\begin{equation}
    b = \sqrt{\frac{\Delta}{|p|}}\,c\,,\quad \nu = \frac{4}{\hbar\kappa\gamma\sqrt{\Delta}}\,{\rm sgn}(p)\,|p|^{3/2}\,,\quad \{b,\nu\}= \frac{2}{\hbar}\,.
    \label{eq:newvariables}
\end{equation}
In these conventions, $b$ and $\nu$ are both dimensionless. $b$ is related to the Hubble rate and $\nu$ corresponds to a signed volume, both measured in Planck units given by $\Delta$; indeed, we observed earlier that the spatial volume grows as $|p|^{3/2}$. Moreover, the holonomy operators are simply defined as $\hat{\mathcal{N}}_{\bar\mu} = \widehat{e^{\frac{{\rm i}}{2}b}}$ and we have
\begin{equation}
  [ \hat\nu, \hat{\mathcal{N}}_{\bar\mu} ] =  \hat{\mathcal{N}}_{\bar\mu}\,,
\end{equation}
i.e., the action of $\hat{\mathcal{N}}_{\bar\mu}$ corresponds to translating the $\nu$ variable by one (Planckian) unit.

In the new variables, the Hamiltonian constraint  (\ref{eq:constraint}) takes the form
\begin{equation}
   -\frac{3}{\kappa\gamma^2\Delta}\,b^2 +\frac{8}{\hbar^2\kappa^2\gamma^2 \Delta} \frac{\pi_\phi^2}{\nu^2}+ \frac{\Lambda}{\kappa} +V(\phi)\approx 0
   \label{eq:newconstraint}
\end{equation}

The task is now to convert this constraint into a well-defined operator. The main step is to replace $b$ by a suitable object constructed from holonomy operators, but further regularisation issues arise because one also needs to represent inverse powers of the volume $\nu$. There is a vast literature on these procedures and the ambiguities they introduce (e.g., \cite{Bojowald:2007ra}); here we will follow the improved dynamics prescription \cite{Banerjee:2011qu,Chiou:2010ne,improve} in its simplified form, the so-called solvable loop quantum cosmology \cite{Ashtekar:2007em}. We will also restrict ourselves to effective equations, obtained from classical regularisation or alternatively from evaluating expectation values in semiclassical states in the quantum theory \cite{Taveras:2008ke}. In practice, this amounts to the simple substitution $b\rightarrow\sin(b)$ in the constraint, which means we replace $b$ by the operator
\begin{equation}
 \widehat{\sin(b)} = \frac{1}{2\im}\left(\widehat{e^{\im b}}-\widehat{e^{-\im b}}\right) = \frac{1}{2\im}\left(\hat{\mathcal{N}}_{2\bar\mu} - \hat{\mathcal{N}}_{-2\bar\mu} \right)\,.
\end{equation}
While this replacement is the simplest implementation of a modified Hamiltonian constraint using input from loop quantum gravity, it is certainly not a unique procedure (see, e.g., \cite{Assanioussi:2018hee} for an alternative model resulting from different choices in the quantisation). 
Effective equations are expected to capture properties of certain semiclassical states and allow for a classical analysis of the modified constraint, which is what we will do in the following.

\section{Classical solutions}
\label{classsol}

It is instructive to first derive classical solutions without including any holonomy corrections. In this case the Hamiltonian constraint is the one derived in (\ref{eq:newconstraint}),
\begin{equation}
   \mathcal{H} \equiv -\frac{3}{\kappa\gamma^2\Delta}\,b^2 +\frac{8}{\hbar^2\kappa^2\gamma^2 \Delta} \frac{\pi_\phi^2}{\nu^2}+ \frac{\Lambda}{\kappa} +V(\phi)\approx 0\,,
\end{equation}
and we recall the somewhat nonstandard choice of Poisson brackets
\begin{equation}
\{ b, \nu \} = \frac{2}{\hbar}\,,\quad \{ \phi,\pi_\phi\} = 1 \,, \quad \{T,\Lambda \}=\kappa\,.
\end{equation}
Using $\mathcal{H}$ as the Hamiltonian for time evolution means that we have fixed the unimodular gauge $N\,\sqrt{q}=\frac{\sqrt{{}^0 q}}{\mathcal{V}}$ or $N=|p|^{-3/2}=\frac{4}{\hbar\kappa\gamma\sqrt{\Delta}|\nu|}$ (recall (\ref{eq:physmetric}) and (\ref{eq:newvariables}) for the relation between $\nu$, $p$ and the spatial metric $q_{ab}$) for now. We will later reparametrise the theory to pass to different choices of time coordinate. In the unimodular gauge, the equations of motion are then given by Hamilton's equations
\begin{align}
    \dot \nu &= \frac{12}{\kappa \hbar\gamma^2\Delta}\, b\,,\qquad
    \dot b = -\frac{32}{\kappa^2 \hbar^3 \gamma^2 \Delta}\,\frac{\pi_\phi^2}{\nu^3}\,,\\
    \dot \phi &= \frac{16}{\kappa^2 \hbar^2 \gamma^2 \Delta}\,\frac{\pi_\phi}{\nu^2}\,, \qquad     \dot \pi_\phi = -V'(\phi)\, .
\end{align}
The equations of motion for $T$ and $\Lambda$ are trivial; we have $\dot{T}=1$ and $\dot\Lambda=0$ as expected.

These equations can be solved analytically for special choices of potential $V(\phi)$. The simplest case is a constant potential $V(\phi)=W={\rm const}$ acting as a shift to the cosmological constant, and we can define $\tilde\Lambda=\Lambda+\kappa\, W$ to simplify the notation. In this case $\pi_\phi$ is a constant of motion. We are particularly interested in the equation of motion for $\nu$. Squaring and eliminating $b^2$ using the constraint, we have
\begin{equation}
\label{eq:classicaluniFried}
    \dot \nu^2 =\frac{48}{\kappa^2\hbar^2\gamma^2 \Delta}\left( \frac{8}{\kappa\hbar^2\gamma^2\Delta}\,\frac{\pi_\phi^2}{\nu^2}+\tilde{\Lambda}\right)\,,
\end{equation}
from which we get the solutions
\begin{align}
    \nu(T) &= \pm \left(\frac{6}{\kappa^3 }\right)^{1/4} \frac{4\sqrt{|\pi_\phi|}}{\hbar\gamma\sqrt{\Delta}}\sqrt{|T-T_0|}\,,\qquad \tilde{\Lambda} = 0\,,\label{eq:vclassT L0}\\
    \nu(T) &= \pm \frac{4\sqrt{3}}{\kappa\hbar\gamma\sqrt{\Delta}}\sqrt{\tilde{\Lambda}(T-T_0)^2-\frac{\kappa\pi_\phi^2}{6\tilde{\Lambda}}}\,,\qquad \tilde{\Lambda}\neq 0\,.\label{eq:vclassT Lnot0}
\end{align}
As the relevant equations are all invariant under $\nu\rightarrow -\nu$ and $b\rightarrow -b$, the sign of $\nu(T)$ is always arbitrary, and in the physical interpretation of the solutions only corresponds to an orientation factor. Below we will often give the solutions for $\nu^2$ instead, suppressing these ambiguities.

(\ref{eq:vclassT Lnot0}) includes the cases of both negative and positive $\tilde\Lambda$. We can see that the domains in which these solutions are defined depends on the sign:  $|T-T_0| \lessgtr \sqrt{\frac{\kappa \pi_\phi^2}{6 \tilde{\Lambda^2}}}$ when $\tilde{\Lambda}\lessgtr 0$. The negative $\tilde\Lambda$ case describes a Universe coming out of the Big Bang, recollapsing and ending in the Big Crunch, whereas positive $\tilde\Lambda$ solutions either expand or contract forever, and only have a single singularity in either past or future. Big Bang/Crunch singularities occur when $|T-T_0| = \sqrt{\frac{\kappa \pi_\phi^2}{6 \tilde{\Lambda^2}}}.$

\begin{figure}[htbp]
    \centering
    \begin{minipage}{0.32\textwidth}
        \centering
        \begin{subfigure}[b]{\textwidth}
            \centering
           \includegraphics[width=\linewidth]{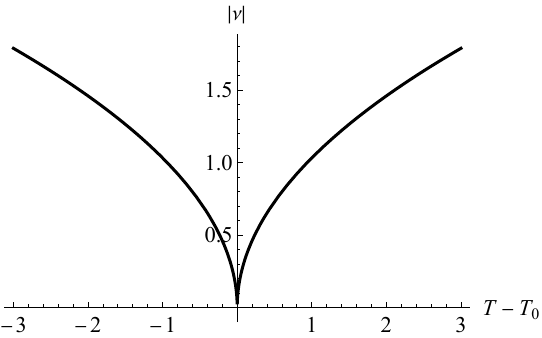}
            \caption{$\tilde{\Lambda}=0$}
        \end{subfigure}
        \vfill
        \begin{subfigure}[b]{\textwidth}
            \centering            \includegraphics[width=\linewidth]{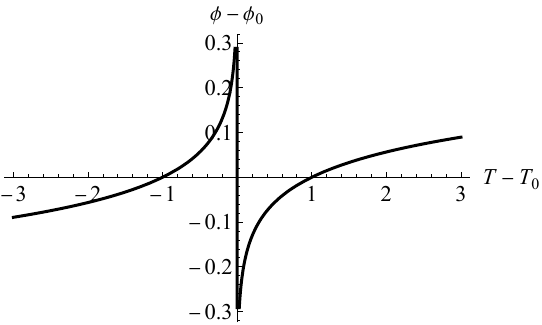}
            \caption{$\tilde{\Lambda}=0$}
        \end{subfigure}
    \end{minipage}%
    \hfill
    \begin{minipage}{0.32\textwidth}
        \begin{subfigure}[b]{\textwidth}
            \centering            \includegraphics[width=\linewidth]{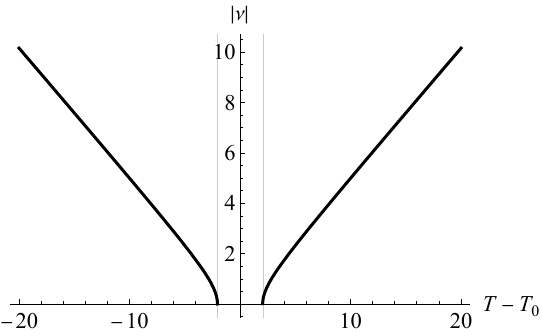}
            \caption{$\tilde{\Lambda}=1$}
        \end{subfigure}
        \vfill
        \begin{subfigure}[b]{\textwidth}
            \centering            \includegraphics[width=\linewidth]{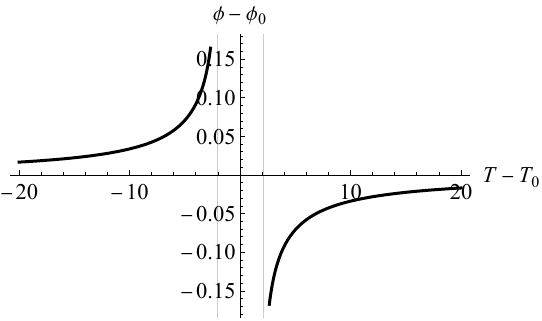}
            \caption{$\tilde{\Lambda}=1$}
        \end{subfigure}
    \end{minipage}%
    \hfill
    \begin{minipage}{0.32\textwidth}
        \begin{subfigure}[b]{\textwidth}
            \centering            \includegraphics[width=\linewidth]{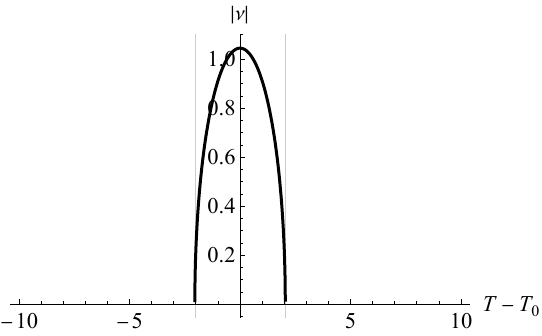}
            \caption{$\tilde{\Lambda}=-1$}
        \end{subfigure}
        \vfill
        \begin{subfigure}[b]{\textwidth}
            \centering            \includegraphics[width=\linewidth]{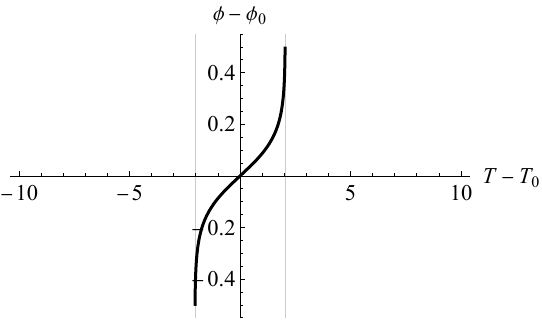}
            \caption{$\tilde{\Lambda}=-1$}
        \end{subfigure}
    \end{minipage}
    \caption{Classical solutions for constant potential. The vertical grey lines indicate $T-T_0 = \pm \sqrt{\frac{\kappa\, \pi_\phi^2}{6 \Lambda^2}}$. We also choose $\pi_\phi = 1$.}
    \label{fig:class const L}
\end{figure}

This leads to different solutions for $\phi(T)$ in each case:
\begin{align}
    \phi(T) &= \frac{{\rm sgn}(\pi_\phi(T-T_0))}{\sqrt{6\kappa}}\log|T-T_0|+\phi_0,\qquad \tilde{\Lambda} = 0,\\
    \phi(T) &= -\sqrt{\frac{2}{3\kappa}}\textrm{arcoth}\left(\frac{\sqrt{6}\tilde\Lambda}{\sqrt{\kappa}\pi_\phi}(T-T_0)\right)+ \phi_0, \qquad \tilde{\Lambda} > 0,\\
    \phi(T) &= -\sqrt{\frac{2}{3\kappa}}\textrm{artanh}\left(\frac{\sqrt{6}\tilde\Lambda}{\sqrt{\kappa}\pi_\phi}(T-T_0)\right)+ \phi_0, \qquad \tilde{\Lambda} < 0.
\end{align}
An example of these solutions for particular choices of the free parameters is plotted in figure \ref{fig:class const L}. In all plots in this and the next section, we choose units in which $\hbar=G=1$ and hence $\kappa=8\pi$. We also choose $\gamma=0.2375$, in agreement with a calculation of entropy in loop quantum gravity \cite{Meissner:2004ju}. The area gap $\Delta$ is fixed by \eqref{eq:improdyn}.

As we noted above, in our setting the choice of unimodular time $T$ as an evolution parameter is most natural, as this would also be the time variable of the quantum theory. However, to find analytical solutions it will later be easier to work in cosmic time $t$ (with $N=1$), so we also give the solutions in cosmic time for completeness here. 

Given that the lapse choice for unimodular gauge was $N=4/(\hbar\kappa\gamma\sqrt{\Delta}|\nu|)$, the Friedmann equation in cosmic time takes the form
\begin{equation}
    \dot \nu^2 = \frac{24}{\kappa\hbar^2\gamma^2\Delta}\,\pi_\phi^2 +3\tilde{\Lambda}\nu^2\,,
\end{equation}
which can be easily solved to give
\begin{align}
    \nu(t) &= \pm\sqrt{\frac{24\pi_\phi^2}{\kappa\hbar^2\gamma^2\Delta}}(t-t_0)\,, \qquad \tilde\Lambda = 0\,,\label{eq:vclasst L0}\\
    \nu(t) &= \pm\sqrt{\frac{8\pi_\phi^2}{\kappa\hbar^2\gamma^2\Delta\,\tilde\Lambda}}\sinh\left[\sqrt{3\tilde\Lambda}(t-t_0)\right]\,, \qquad \tilde\Lambda \neq 0\,.\label{eq:vclasst Lnot0}
\end{align}
Note that when $\tilde{\Lambda} < 0$ the second solution is still real, as for real $x$ we have $\sinh(\im x)/(\im x)=\sin(x)/x$ and the solution turns into a sine including $|\tilde\Lambda|$. These solutions remain well-defined and differentiable at  singularities $\nu\rightarrow 0$, and connect what might be seen as different cosmological solutions.

The equation of motion for $\phi$ now becomes
\begin{equation}
 \dot \phi = \frac{4}{\kappa \hbar \gamma \sqrt{\Delta}}\,\frac{\pi_\phi}{|\nu|}\,,
\end{equation}
and it is solved by
\begin{align}
    \phi(t) &= {\rm sgn}(\pi_\phi(t-t_0)) \sqrt{\frac{2}{3\kappa}}\log|t-t_0|+\phi_0, \qquad\qquad\qquad\qquad\qquad \tilde\Lambda=0,\\
    \phi(t) &= {\rm sgn}(\pi_\phi(t-t_0))\sqrt{\frac{2}{3\kappa}}\log\left| \tanh\left(\frac{\sqrt{3\tilde\Lambda}}{2}(t-t_0)\right) \right|+\phi_0\,, \qquad\qquad\qquad \tilde\Lambda > 0\,,\\
    \phi(t) &= \textrm{sgn}\left(\pi_\phi\,\sin(\sqrt{3|\tilde\Lambda|}\,(t-t_0))\right)\sqrt{\frac{2}{3\kappa}}\log\left|\tan\left(\frac{\sqrt{3|\tilde\Lambda|}}{2}(t-t_0)\right) \right|+\phi_0\,, \quad \tilde\Lambda < 0\,.
\end{align}

Finally, we might also be interested in writing the solutions in terms of the scalar field $\phi$, which appears as the preferred clock in standard loop quantum cosmology. In this case the Friedmann equation takes the form
\begin{equation}
 \dot \nu^2 =\frac{3\kappa}{2}\left( \frac{\kappa h^2 \gamma^2\Delta}{8\pi_\phi^2} \tilde\Lambda\,\nu^4+\nu^2\right)\,,
\end{equation}
which admits the following solutions:
\begin{align}
    \nu(\phi) &= \pm \sqrt{\frac{24\pi_\phi^2}{\kappa\hbar^2}}e^{\pm\sqrt{\frac{3\kappa}{2}}(\phi-\phi_0)}\,,\qquad \tilde\Lambda = 0\,,\label{eq:vclassphi L0}\\
    \nu(\phi) &= \pm \sqrt{\frac{8\pi_\phi^2}{\kappa\hbar^2\gamma^2\Delta\,\tilde\Lambda}}\ \frac{1}{\left\vert\sinh\left(\sqrt{\frac{3\kappa}{2}}(\phi-\phi_0)\right)\right\vert}\,,\qquad \tilde{\Lambda}>0\,,\label{eq:vclassphi Lpos}\\
    \nu(\phi) &= \pm \sqrt{\frac{8\pi_\phi^2}{\kappa\hbar^2\gamma^2\Delta\,|\tilde\Lambda|}}\ \frac{1}{\cosh\left(\sqrt{\frac{3\kappa}{2}}(\phi-\phi_0)\right)}\,,\qquad \tilde{\Lambda}<0\,.\label{eq:vclassphi Lneg}
\end{align}
In (\ref{eq:vclassphi L0}) there are two independent signs to choose, as the scalar field already takes all possible real values within only one of the two (expanding or contracting) solution branches. Hence, the clock $\phi$ can only cover one of these two.

One can extend this analysis to other examples of potentials $V(\phi)$ where analytical solutions are known. For instance, it is well-known that solutions corresponding to a general power law of the volume in cosmic time can be engineered by choosing an exponential potential in $\phi$ (see, e.g., \cite{Ratra:1987rm}). In the more general case, the Friedmann equation in cosmic time is
\begin{equation}
    \dot \nu^2 = \frac{24}{\kappa\hbar^2\gamma^2\Delta}\,\pi_\phi^2 +3\nu^2\left(\Lambda+\kappa V(\phi)\right)\,.
\end{equation}
Let us assume that $\Lambda=0$ and 
\begin{equation}
V(\phi)=V_0\exp\left(-\sqrt{\frac{2\kappa}{p}}\phi\right)
\end{equation}
for some parameter $p>0$, where one is usually interested in $p>1$. One can then check that the following is a solution to the dynamical equations and the Friedmann constraint:
\begin{align}
   \nu(t) &= \nu_0 |t-t_0|^{3p}\,,
 \\ \phi(t) & =  \sqrt{\frac{2p}{\kappa}} \log\left(\sqrt{\frac{\kappa\,V_0}{p(3p-1)}}|t-t_0|\right)\,.
\end{align}
Since there is no conserved momentum $\pi_\phi$ in this case, the initial condition parameter $\nu_0$ is not related to any conserved quantity and can be chosen arbitrarily. One can also express $\nu$ in terms of the $\phi$ clock:
\begin{equation}
 \nu(\phi) = \nu_0 \left(\frac{p(3p-1)}{\kappa V_0}\right)^{\frac{3p}{2}}\exp\left(3\sqrt{\frac{\kappa p}{2}}\phi\right)\,.
\end{equation}

These solutions are in general fine-tuned and obtaining the general solution would be substantially more complicated.

\section{Analytical solutions in effective loop quantum cosmology}
\label{effectivesol}

In the simplest setting for loop quantum cosmology that we are interested in here, we now include holonomy corrections into the Hamiltonian constraint by replacing $b\rightarrow\sin(b)$. This yields
\begin{equation}
    \mathcal{H} = -\frac{3}{\kappa\gamma^2\Delta} \sin^2(b)+\frac{8}{\hbar^2 \kappa^2  \gamma^2 \Delta} \frac{\pi_\phi^2}{\nu^2}+ \frac{\Lambda}{\kappa} +V(\phi) \approx 0\,,
\end{equation}
from which we obtain
\begin{equation}\label{eq:sinb_constraint}
    \sin^2(b) = \frac{8}{3 \hbar^2\kappa } \frac{\pi_\phi^2}{\nu^2}+\frac{\gamma^2\Delta}{3} \left(\Lambda + \kappa V(\phi)\right)
\end{equation}
and the equation of motion for $\nu$ in the original unimodular gauge,
\begin{equation}
    \dot \nu = \frac{12}{\kappa \hbar\gamma^2\Delta}\, \sin(b)\cos(b)\,.
\end{equation}
The other dynamical equations are unchanged. If we choose a constant potential $V(\phi)=W$, then $\pi_\phi$ is again a constant of motion. By squaring the equation of motion for $\nu$ and eliminating $b$ using the constraint, we obtain a modified Friedmann equation (again with $\tilde\Lambda=\Lambda+\kappa W$)
\begin{align}
    \dot \nu^2 &= \left(\frac{12}{\kappa \hbar\gamma^2\Delta}\right)^2\ \left(\frac{\gamma^2\Delta}{3}\tilde{\Lambda}\left(1-\frac{\gamma^2\Delta}{3}\tilde{\Lambda}\right)+\frac{8\pi_\phi^2}{3 \hbar^2 \kappa }\left(1-\frac{2\gamma^2 \Delta}{3}\tilde{\Lambda}\right) \frac{1}{\nu^2}-\left(\frac{8\pi_\phi^2}{3 \kappa \hbar^2}\right)^2\frac{1}{\nu^4}\right)\nonumber\\
    &\equiv \left(\frac{12}{\kappa \hbar\gamma^2\Delta}\right)^2 \left(A+\frac{B}{\nu^2}-\frac{C^2}{\nu^4}\right)\,,
\end{align}
where we define the constants
\begin{equation}
    A = \frac{\gamma^2\Delta}{3}\tilde{\Lambda}\left(1-\frac{\gamma^2\Delta}{3}\tilde{\Lambda}\right)\,,\quad B= C\left(1-\frac{2\gamma^2\Delta}{3}\tilde{\Lambda}\right)\,,\quad C = \frac{8\pi_\phi^2}{3 \kappa \hbar^2}\ . 
\end{equation}
Comparison with (\ref{eq:classicaluniFried}) shows that the classical limit would correspond to taking $A=\frac{\gamma^2\Delta}{3}\tilde\Lambda$ and $B=C$ while setting the repulsive $1/\nu^4$ term to zero. For a sub-Planckian cosmological constant and potential, $|\tilde\Lambda| \ll 1/(\gamma^2\Delta)$, the corrections for $A$ and $B$ would be very small; the repulsive  $1/\nu^4$ term is then the main result of including holonomy corrections, as it dominates at very small $\nu$, leading to singularity resolution by a bounce. In practical terms, solving the Friedmann equation is now much more difficult because of the presence of this extra term.

However, even without computing explicit solutions, one can already infer some of the behaviour of $|\nu(T)|$ from requiring that 
\begin{equation}
    \dot \nu^2 \geq 0 \quad \Leftrightarrow  \quad A\nu^4+B\nu^2-C^2\geq0\,.
\end{equation}
The roots of the polynomial are 
\begin{equation}
    \nu^2_c=\frac{-B\pm\sqrt{B^2+4A\,C^2}}{2A} =\frac{-B\pm C}{2A}= \left\{ \frac{8\pi_\phi^2}{\kappa\hbar^2(3-\gamma^2\Delta\tilde{\Lambda})} \,, \, -\frac{8\pi_\phi^2}{\kappa\hbar^2\gamma^2\Delta \tilde{\Lambda}} \right\}\,.
\end{equation}
Then, by looking at the possible signs for $A$ and $B$, we see that:
\begin{itemize}
    \item If $A>0$, which means $0 < \tilde\Lambda < \frac{3}{\gamma^2\Delta}$, there are only solutions for $\nu^2\geq \frac{8\pi_\phi^2}{\kappa\hbar^2(3-\gamma^2\Delta\tilde{\Lambda})}>0$. Since $\nu^2$ is bounded away from zero, these solutions correspond to a bounce interpolating between classical contracting and expanding solutions. 
    \item If $A<0$, solutions are only possible for  $B>0$. This means that any $\tilde\Lambda<0$ is possible whereas $\tilde\Lambda>\frac{3}{\gamma^2\Delta}$ has no solutions. Negative $\tilde\Lambda$ leads to solutions with $\frac{8\pi_\phi^2}{\kappa\hbar^2(3+\gamma^2\Delta|\tilde{\Lambda}|)}\leq \nu^2 \leq \frac{8\pi_\phi^2}{\kappa\hbar^2\gamma^2\Delta |\tilde{\Lambda}|}$, which correspond to a cyclic Universe with successive bounces and recollapses.
    \item If $A=0$, there are only solutions for $B > 0$. This means that $\tilde\Lambda=0$ is possible, but $\tilde\Lambda=\frac{3}{\gamma^2\Delta}$ has no solutions. The case $\tilde\Lambda=0$ corresponds to a bounce with $\nu^2 \geq \frac{8\pi_\phi^2}{3\kappa\hbar^2}$. 
\end{itemize}
We conclude that $\tilde{\Lambda} < \frac{3}{\gamma^2\Delta}$, which corresponds to a Planck-scale upper bound on the (effective) cosmological constant, and that in all cases the singularity is resolved as there is always a minimum $|\nu|$ that is strictly greater than zero. Writing the equations of motion in other time coordinates allows us to provide analytical solutions for all the possible cases.

In cosmic time $t$ (with $N=1$) the Hamiltonian is
\begin{equation}
    \mathcal{H} = -\frac{3\hbar}{4\gamma\sqrt{\Delta}} \sin^2 (b) |\nu|+\frac{2}{\kappa \hbar \gamma\sqrt{\Delta}} \frac{\pi_\phi^2}{|\nu|}+\frac{\hbar\gamma\sqrt{\Delta}}{4} |\nu| \tilde{\Lambda}\,,
    \label{cosmicLQCHamilt}
\end{equation}
from which we obtain the equations of motion
\begin{equation}
\dot\nu = \frac{3}{\gamma\sqrt{\Delta}}|\nu|\sin(b) \cos(b)\,,\quad \dot\phi = \frac{4}{\kappa\hbar\gamma\sqrt{\Delta}}\frac{\pi_\phi}{|\nu|}\,, \quad\dot{T} = \frac{\hbar\kappa\gamma\sqrt{\Delta}}{4} |\nu| 
\label{eq:cosmiceom}
\end{equation}
and the Friedmann equation
\begin{equation}
\dot{\nu}^2 = \frac{9}{\gamma^2\Delta}\left(A\nu^2+B-C^2 \frac{1}{\nu^2}\right)
\label{newproperfried}
\end{equation}
with $A$, $B$ and $C$ as defined above. Again, there are solutions only for $\tilde{\Lambda}<\frac{3}{\gamma^2\Delta}$. With this choice of time we find analytical solutions, now written in terms of $\nu^2(t)$ for simplicity:
\begin{align}
    \nu^2(t) &= \frac{8\pi_\phi^2}{3\kappa\hbar^2}\left(1+\frac{9}{\gamma^2 \Delta}(t-t_0)^2\right)\,, \qquad \tilde{\Lambda}=0\,,\label{eq:vefft L0}\\
    \nu^2(t) &=  \frac{C}{2A}\cosh\left(2\sqrt{\tilde{\Lambda}(3-\gamma^2\Delta\tilde{\Lambda})}\,(t-t_0)\right)-\frac{B}{2 A}\nonumber
    \\ & = \frac{4\pi_\phi^2\cosh\left(2\sqrt{\tilde{\Lambda}(3-\gamma^2\Delta\tilde{\Lambda})}\,(t-t_0)\right)}{\hbar^2\kappa\gamma^2\Delta\tilde\Lambda(1-\frac{\gamma^2\Delta\tilde\Lambda}{3})}-\frac{4\pi_\phi^2(3-2\gamma^2\Delta\tilde\Lambda)}{\hbar^2\kappa\gamma^2\Delta\tilde\Lambda(3-\gamma^2\Delta\tilde\Lambda)}\,,\quad \tilde{\Lambda} \neq 0\,,\quad \tilde{\Lambda}<\frac{3}{\gamma^2\Delta}\,. \label{eq:nu(t) Lambda}
\end{align}
For $\tilde\Lambda<0$, the argument of the square root is negative, so that we have an oscillatory solution with $\cosh(\im x)=\cos(x)$ for real $x$. Since in this case $B>C>0$, we can see that the solution remains bounded away from zero due to the positive constant contribution $-\frac{B}{2A}$.

The Friedmann equation (\ref{newproperfried}) can also be brought into a more familiar form by reintroducing a scale factor via $|\nu|=\frac{4}{\hbar \kappa \gamma \sqrt{\Delta}} a^3$. Then the Friedmann equation for $a$ is
\begin{equation}
\left(\frac{\dot{a}}{a}\right)^2 = \frac{\kappa}{3}\rho\left(1-\frac{\rho}{\rho_c}\right)   
\label{standardeffFr}
\end{equation}
with
\begin{equation}
\rho = \frac{\pi_\phi^2}{2a^6}+\frac{\tilde\Lambda}{\kappa}\,,\qquad \rho_c = \frac{3}{\gamma^2\Delta\kappa}\,.
\end{equation}
In this form it is evident that $\rho_c$ is a maximal (critical) value for the energy density at which a bounce occurs. This form of the effective Friedmann equation is often the starting point for phenomenological studies in loop quantum cosmology (see, e.g., \cite{Bhardwaj:2018omt,Paul:2025qtj} for studies of inflation in this context).

With an explicit solution for $\nu(t)$ at hand, solutions for $\phi(t)$ and $T(t)$ can now be obtained by integrating (\ref{eq:cosmiceom}). We find
\begin{align}
\phi(t)&={\rm sgn}(\pi_\phi)\sqrt{\frac{2}{3\kappa}}\,\textrm{arcsinh}\left[\frac{3}{\gamma\sqrt{\Delta}}(t-t_0)\right]+\phi_0\,,\qquad \tilde\Lambda=0\,,\label{eq:phiteff L0}\\
    \phi(t)&=-\im\,{\rm sgn}(\pi_\phi)\sqrt{\frac{2}{\kappa\gamma^2\Delta\tilde{\Lambda}}}\ F\left(\im\sqrt{\tilde\Lambda(3-\gamma^2\Delta\tilde\Lambda)}(t-t_0) \left \vert \frac{3}{\gamma^2\Delta\tilde{\Lambda}} \right. \right)+\phi_0\,,\qquad 0<\tilde{\Lambda}<\frac{3}{\gamma^2\Delta}\,,\label{eq:phiteff Lpos}\\
    \phi(t)&={\rm sgn}(\pi_\phi)\sqrt{\frac{2}{\kappa\gamma^2\Delta|\tilde{\Lambda}|}}\  F\left(\sqrt{|\tilde\Lambda|(3-\gamma^2\Delta\tilde\Lambda)}(t-t_0) \left \vert \frac{3}{\gamma^2\Delta\tilde{\Lambda}} \right. \right)+\phi_0\,,\qquad \tilde{\Lambda}<0\,,\label{eq:phiteff Lneg}
\end{align}
where $F(\phi\vert m)=\int_0^\phi {\rm d}\theta/\sqrt{1-m\sin^2\theta}$ is the incomplete elliptic integral of the first kind. Similarly,
\begin{align}
    T(t) &= \frac{\sqrt{\kappa}|\pi_\phi|}{2\sqrt{6}}
    \left[(t-t_0)\sqrt{\gamma^2\Delta+9(t-t_0)^2}-\frac{\gamma^2\Delta}{3}\log\left(\sqrt{1+\frac{9}{\gamma^2\Delta}(t-t_0)^2}-\frac{3}{\gamma\sqrt{\Delta}}(t-t_0)\right)\right]+T_0\label{eq:Tteff L0}
    \end{align}
    for $\tilde\Lambda=0$ and
    \begin{align}
    T(t) &= -\im\frac{|\pi_\phi|}{(3-\gamma^2\Delta\tilde\Lambda)}\sqrt{\frac{\gamma^2\Delta\kappa}{2\tilde{\Lambda}}}\ E\left(\im\sqrt{\tilde\Lambda(3-\gamma^2\Delta\tilde\Lambda)}\ (t-t_0)\, \left \vert \frac{3}{\gamma^2 \Delta\tilde{\Lambda}}\right. \right)+T_0\,,\quad 0<\tilde{\Lambda}<\frac{3}{\gamma^2\Delta}\,,\label{eq:Tteff Lpos}\\
    T(t) &= \frac{|\pi_\phi|}{(3-\gamma^2\Delta\tilde\Lambda)}\sqrt{\frac{\gamma^2\Delta\kappa}{2|\tilde{\Lambda}|}}\ E\left(\sqrt{|\tilde\Lambda|(3-\gamma^2\Delta\tilde\Lambda)}\ (t-t_0)\, \left \vert \frac{3}{\gamma^2 \Delta \tilde{\Lambda}}\right. \right)+T_0\,,\qquad\tilde\Lambda<0\,,\label{eq:Tteff Lneg}
\end{align}
where $E(\phi\vert m)=\int_0^\phi\sqrt{1-m\sin^2\theta}\,d\theta$ is the incomplete elliptic integral of the second kind.  Again, from their definition one can check that the elliptic integrals in these expressions are always real ($\tilde\Lambda<0$) or purely imaginary ($\tilde\Lambda>0$), meaning we get real-valued solutions for $\phi(t)$ and $T(t)$.

\begin{figure}[htbp]
    \centering
    \begin{minipage}{0.32\textwidth}
        \centering
        \begin{subfigure}[b]{\textwidth}
            \centering
           \includegraphics[width=\linewidth]{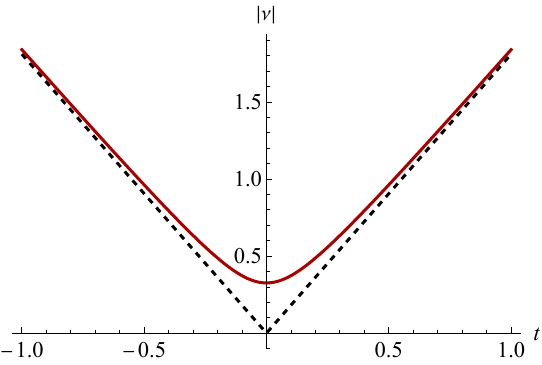}
            \caption{$\tilde{\Lambda}=0$: $t_0=0$ in both \eqref{eq:vclasst L0} and \eqref{eq:vefft L0}.}
            \end{subfigure}
        \vfill
        \begin{subfigure}[b]{\textwidth}
            \centering            \includegraphics[width=\linewidth]{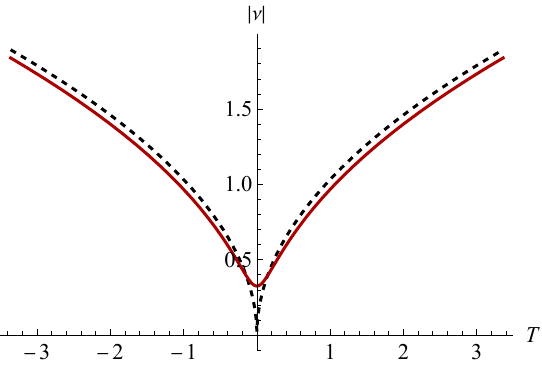}
            \caption{$\tilde{\Lambda}=0$: $T_0=0$ in \eqref{eq:vclassT L0} and \eqref{eq:Tteff L0}.}
        \end{subfigure}
        \vfill
        \begin{subfigure}[b]{\textwidth}
            \centering            \includegraphics[width=\linewidth]{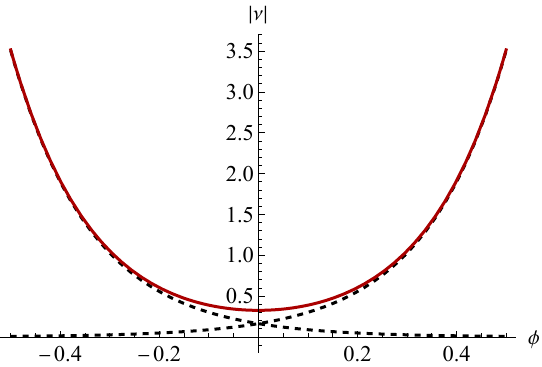}
            \caption{$\tilde{\Lambda}=0$: $\phi_0 = \pm\sqrt{\frac{2}{3\kappa}}\log (6)$ in \eqref{eq:vclassphi L0} and $\phi_0=0$ in \eqref{eq:veffphi L0}.}
        \end{subfigure}
   \end{minipage}%
     \hfill
    \begin{minipage}{0.32\textwidth}
        \centering
        \begin{subfigure}[b]{\textwidth}
            \centering            \includegraphics[width=\linewidth]{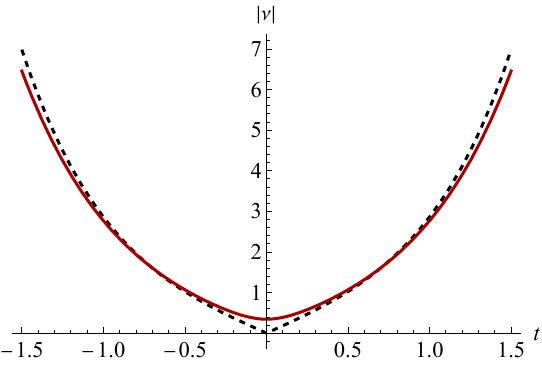}
            \caption{$\tilde{\Lambda}=1$: $t_0=0$ in both \eqref{eq:vclasst Lnot0} and \eqref{eq:nu(t) Lambda}.}
        \end{subfigure}
        \vfill
        \begin{subfigure}[b]{\textwidth}
            \centering            \includegraphics[width=\linewidth]{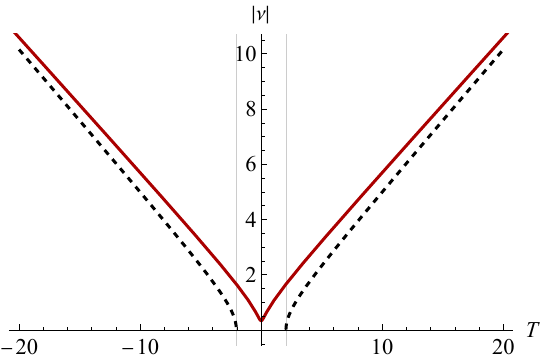}
            \caption{$\tilde{\Lambda}=1$: $T_0=0$ in \eqref{eq:vclassT Lnot0} and \eqref{eq:Tteff Lpos}.}
        \end{subfigure}
        \vfill
    \begin{subfigure}[b]{\textwidth}
            \centering            \includegraphics[width=\linewidth]{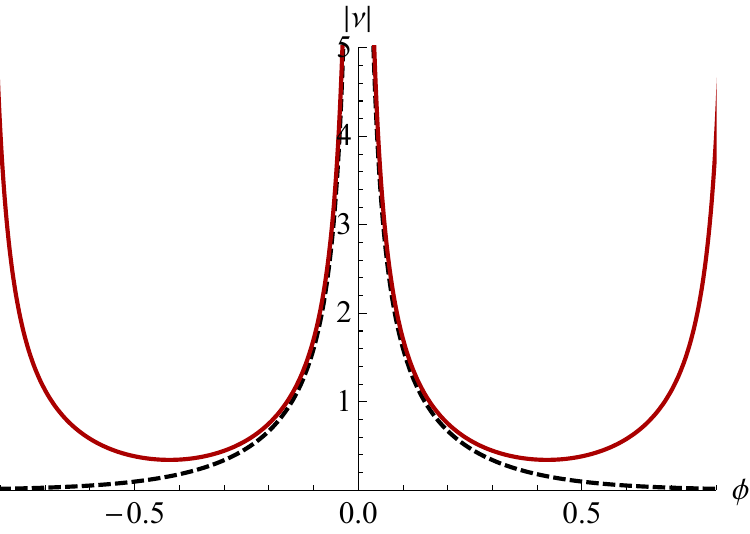}
            \caption{$\tilde{\Lambda}=1$: $\phi_0=0$ in \eqref{eq:vclassphi Lpos}, and $\phi_0=\frac{\pm{\rm sgn}(\pi_\phi)}{\gamma\sqrt{\Delta}}\sqrt{\frac{2}{\kappa\tilde{\Lambda}}}\ K\left(1-\frac{3}{\gamma^2\Delta\tilde\Lambda}\right)$ in \eqref{eq:phiteff Lpos}.}
        \end{subfigure}    \end{minipage}%
    \hfill
    \begin{minipage}{0.32\textwidth}
        \centering
\begin{subfigure}[b]{\textwidth}
            \centering            \includegraphics[width=\linewidth]{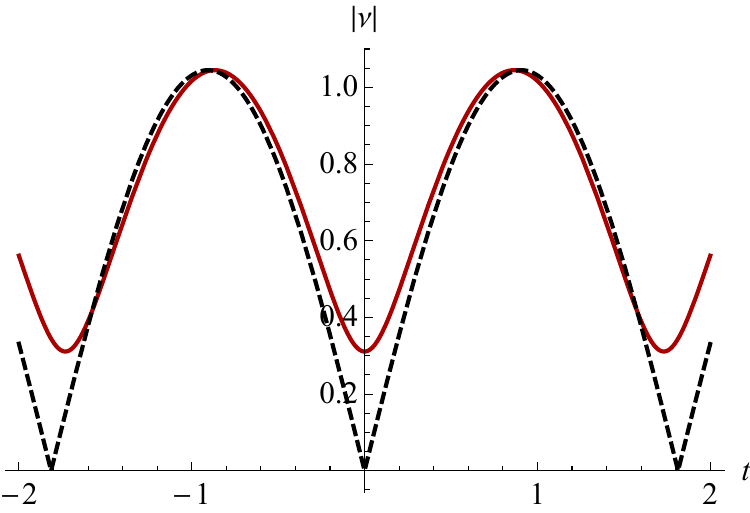}
            \caption{$\tilde{\Lambda}=-1$: $t_0=0$ in both \eqref{eq:vclasst Lnot0} and  \eqref{eq:nu(t) Lambda}.}
        \end{subfigure}
        \vfill
        \begin{subfigure}[b]{\textwidth}
            \centering            \includegraphics[width=\linewidth]{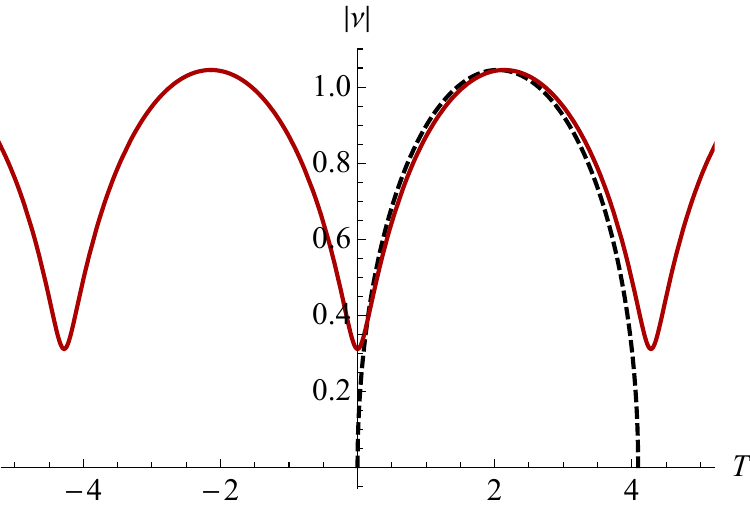}
            \caption{$\tilde{\Lambda}=-1$: $T_0=\sqrt{\frac{\kappa\pi_\phi^2}{6\tilde{\Lambda}^2}}$ in \eqref{eq:vclassT Lnot0} and $T_0=0$ in \eqref{eq:Tteff Lneg}.}
        \end{subfigure}
        \vfill
       \begin{subfigure}[b]{\textwidth}
            \centering            \includegraphics[width=\linewidth]{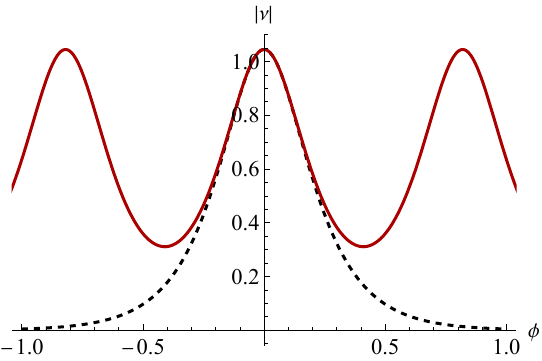}
            \caption{$\tilde{\Lambda}=-1$: $\phi_0=0$ in \eqref{eq:vclassphi Lneg}, and $\phi_0=-\frac{{\rm sgn}(\pi_\phi)}{\gamma\sqrt{\Delta}}\sqrt{\frac{2}{\kappa|\tilde{\Lambda}|}}\  K\left(\frac{3}{\gamma^2\Delta\tilde{\Lambda}}  \right)$  in  \eqref{eq:phiteff Lneg}.}
        \end{subfigure}
    \end{minipage}
    \caption{Classical (black, dashed) vs effective loop quantum cosmology (red, solid) trajectories for constant potential in cosmic (top row), unimodular (middle row) and scalar field time (bottom row).}
    \label{fig:class eff const L}
\end{figure}

We can use these solutions for parametric plots of $|\nu|$ versus one of our relational clocks, either $\phi$ or $T$. These parametric plots, together with plots of the original solutions in proper time, can be found in Figure \ref{fig:class eff const L}. The parametric plots represent relational observables $|\nu|(\phi=\phi_0)$ or $|\nu|(T=T_0)$, the type of observables one can consider in loop quantum cosmology. In these plots, we have chosen the integration constants $\phi_0$ and $T_0$ to allow for a clear comparison of the effective loop quantum cosmology solutions with corresponding classical solutions, where by corresponding we mean that singularities or divergences in classical solutions happen at the same time as bounces in effective loop quantum cosmology.

For $\tilde\Lambda \geq 0$, the classical trajectories \eqref{eq:vclassT L0} and \eqref{eq:vclassT Lnot0} are singular at $T=T_0$, and we set $T_0=0$. Then, in the effective dynamics we want to make sure the minimum of $|\nu(t(T))|$ also happens at $T=0$. The loop quantum cosmology solutions \eqref{eq:vefft L0} and \eqref{eq:nu(t) Lambda} reach their minimum of $|\nu(t)|$ at $t=t_0$, which is indeed where $T=T_0$ in \eqref{eq:Tteff L0} and \eqref{eq:Tteff Lpos} also has a minimum. Hence we set $T_0=0$ there as well.

For $\tilde\Lambda <0$, in the classical trajectory \eqref{eq:vclassT Lnot0} the singularities appear when $|T-T_0| =  \sqrt{\frac{\kappa\pi_\phi^2}{6\tilde\Lambda^2}}$. We choose to fix $T_0= \sqrt{\frac{\kappa\pi_\phi^2}{6\tilde\Lambda^2}}$ so that the Big Bang is at $T=0$, as in the other cases. In the effective dynamics \eqref{eq:nu(t) Lambda}, keeping in mind that in this case $A<0$, we find that the minima are at $t-t_0=n\pi/(\sqrt{|\tilde{\Lambda}|(3-\gamma^2\Delta\tilde{\Lambda})}),\ n\in \mathbb{Z}$. In this case, choosing $T_0=0$ in \eqref{eq:Tteff Lneg} ensures that the effective loop quantum cosmology solution has a minimum at the classical Big Bang $T=0$.

For $\tilde\Lambda = 0$, we can combine \eqref{eq:vefft L0} and \eqref{eq:phiteff L0} to find a well-known explicit expression for $\nu(\phi)$,
\begin{equation}
\nu(\phi) = \pm\sqrt{\frac{8\pi_\phi^2}{3\kappa\hbar^2}}\cosh\left[\sqrt{\frac{3\kappa}{2}}(\phi-\phi_0)\right]\,,\qquad \tilde{\Lambda}=0\,.\label{eq:veffphi L0}
\end{equation}
While this solution has a bounce for $\phi=\phi_0$, the corresponding classical solutions are either purely expanding or purely contracting, with the singularity pushed out to $\phi\rightarrow\pm\infty$. To compare classical and effective loop quantum cosmology solutions, for this case we decide to set $\phi_0=0$ in \eqref{eq:veffphi L0}, and choose $\phi_0$ in \eqref{eq:vclassphi L0} to ensure that the classical solution agrees with the approximate form $\nu(\phi)\approx \pm\sqrt{\frac{2\pi_\phi^2}{3\kappa\hbar^2}}\,e^{\sqrt{\frac{3\kappa}{2}}|\phi|}$ of the effective loop quantum cosmology solution away from the bounce.

In the cases $\tilde\Lambda\neq 0$, one may give $\nu(\phi)$ in terms of Jacobi elliptic functions \cite{abramowitz}. For example, for $\tilde\Lambda<0$, from \eqref{eq:phiteff Lneg} we have
\begin{equation}
\sqrt{\frac{\kappa\gamma^2\Delta|\tilde{\Lambda}|}{2}}\,{\rm sgn}(\pi_\phi)\left(\phi - \phi_0 \right) =   F\left(\sqrt{|\tilde\Lambda|(3-\gamma^2\Delta\tilde\Lambda)}(t-t_0) \left \vert \frac{3}{\gamma^2\Delta\tilde{\Lambda}} \right. \right)
\end{equation}
which is equivalent to
\begin{equation}
\sin\left(\sqrt{|\tilde\Lambda|(3-\gamma^2\Delta\tilde\Lambda)}(t-t_0)\right) = {\rm sn}\left( \sqrt{\frac{\kappa\gamma^2\Delta|\tilde{\Lambda}|}{2}}\,{\rm sgn}(\pi_\phi)\left(\phi - \phi_0 \right) \left\vert \frac{3}{\gamma^2\Delta\tilde{\Lambda}} \right. \right)\,,
\end{equation}
where sn is one of the Jacobi elliptic functions. We would then obtain
\begin{equation}
\nu^2(\phi) = \frac{8\pi_\phi^2}{\hbar^2\kappa(3-\gamma^2\Delta\tilde\Lambda)}-\frac{8\pi_\phi^2\,{\rm sn}^2 \left( \sqrt{\frac{\kappa\gamma^2\Delta|\tilde{\Lambda}|}{2}}\,{\rm sgn}(\pi_\phi)\left(\phi - \phi_0 \right) \left\vert \frac{3}{\gamma^2\Delta\tilde{\Lambda}} \right. \right)}{\hbar^2\kappa\gamma^2\Delta\tilde\Lambda(1-\frac{\gamma^2\Delta\tilde\Lambda}{3})}\,,
\end{equation}
which is equivalent to the implicit definition via parametric plots. The analytic properties of the Jacobi elliptic functions are studied in a very similar context (namely, as the general solutions of a Friedmann equation) in \cite{Boyle:2021jej}.

In this $\tilde\Lambda<0$ case, the comparison between classical and effective loop quantum cosmology solutions is made by aligning a maximum of the effective loop quantum cosmology solution with the maximum in the classical solution. With $\phi_0=0$ in \eqref{eq:vclassphi Lneg}, the maximum is at $\phi=0$; hence, we need to ensure that $\phi=0$ in the effective loop quantum cosmology solution for a $t$ such that a maximum is reached, e.g., $t=t_0+\pi/(2\sqrt{|\tilde{\Lambda}|(3-\gamma^2\Delta\tilde{\Lambda})})$. This can be done by suitable choice of $\phi_0$:
\begin{equation}
  \phi_0 =-{\rm sgn}(\pi_\phi)\sqrt{\frac{2}{\kappa\gamma^2\Delta|\tilde{\Lambda}|}}\  F\left(\frac{\pi}{2} \left \vert \frac{3}{\gamma^2\Delta\tilde{\Lambda}} \right. \right)  =-{\rm sgn}(\pi_\phi)\sqrt{\frac{2}{\kappa\gamma^2\Delta|\tilde{\Lambda}|}}\  K\left(\frac{3}{\gamma^2\Delta\tilde{\Lambda}}  \right)\,,
\end{equation}
where $K(x)$ is now the complete elliptic integral of the first kind.

For $\tilde\Lambda>0$, the effective loop quantum cosmology solution \eqref{eq:phiteff Lpos} for $\phi(t)$ has the particular property that $\phi$ asymptotes to finite values:
\begin{equation}
 \phi(t)\longrightarrow \pm\,{\rm sgn}(\pi_\phi)\sqrt{\frac{2}{\kappa\gamma^2\Delta\tilde{\Lambda}}}\ K\left(1-\frac{3}{\gamma^2\Delta\tilde\Lambda}\right)+\phi_0\,, \quad t\rightarrow \pm \infty\,.
\end{equation}
Depending on the value of $\tilde\Lambda$, the complete elliptic integral takes a value between $\frac{\pi}{2}$ and $\pi$. Hence, the entire evolution of the Universe, through contraction from infinity, bounce and expansion back to infinite volume, is contained within a finite amount of $\phi$ time, with $\nu(\phi)$ diverging at finite $\phi$.

In the effectively classical formalism we are using here, such divergences can arise both in classical and in effective loop quantum cosmology solutions. On the other hand, in a quantum theory where $\phi$ is the time evolution parameter, unitarity in $\phi$ would require reflecting boundary conditions at large volume, leading to large quantum corrections when the classical solution diverges and a continuation of the quantum solution beyond these points. This behaviour was observed for loop quantum cosmology in \cite{Pawlowski:2011zf} and for Wheeler--DeWitt quantum cosmology in \cite{Gielen:2021igw}. In our work, we suggest the use of $T$ as the fundamental clock variable in a quantum theory, and so we would not necessarily require time evolution in $\phi$ to be unitary.

To ensure that divergences in $\nu(\phi)$ for $\tilde\Lambda>0$ appear at $\phi=0$ in both classical and effective loop quantum cosmology solutions, we can choose $\phi_0=0$ in \eqref{eq:vclassphi Lpos} and $\phi_0=\pm\,{\rm sgn}(\pi_\phi)\sqrt{\frac{2}{\kappa\gamma^2\Delta\tilde{\Lambda}}}\ K\left(1-\frac{3}{\gamma^2\Delta\tilde\Lambda}\right)$ in \eqref{eq:phiteff Lpos}.

\section{Numerical solutions for inflationary models}
\label{numericalsol}

In the interest of studying inflationary models, we now turn our attention to non-constant potentials $V(\phi)$. In this situation, one does not expect to have analytical solutions and so the dynamics for such models are usually studied numerically. Some forms of potential that have been studied are a quadratic potential \cite{Ashtekar:2011rm,Agullo:2013ai}, Starobinsky inflation \cite{Bonga:2015kaa,Iteanu:2022zha}, monodromy potentials \cite{Sharma:2018vnv}, $\alpha$-attractors \cite{Shahalam:2018rby,Shahalam:2019mpw} and polynomial chaotic potentials in ekpyrotic scenarios \cite{Brown:2025hcb,Frion:2025cyd}. Our work here extends some of this past work by using a wider range of initial conditions, and by focusing on the choice of unimodular time as an evolution parameter.

Working in cosmic time again, the effective constraint is the one given in (\ref{cosmicLQCHamilt}): 
\begin{equation}
    \mathcal{H} = -\frac{3\hbar}{4\gamma\sqrt{\Delta}} \sin^2 (b) |\nu|+\frac{2}{\kappa \hbar \gamma\sqrt{\Delta}} \frac{\pi_\phi^2}{|\nu|}+\frac{\hbar\gamma\sqrt{\Delta}}{4} |\nu| \tilde\Lambda(\phi)\,,
\end{equation}
where $\tilde\Lambda(\phi)=\Lambda+\kappa V(\phi)$ is now a function of $\phi$ through the contribution of the nontrivial $V(\phi)$.

The equations of motion
\begin{equation}
\dot\nu = \frac{3}{\gamma\sqrt{\Delta}}|\nu|\sin(b) \cos(b)\,,\quad \dot\phi = \frac{4}{\kappa\hbar\gamma\sqrt{\Delta}}\frac{\pi_\phi}{|\nu|}\,, \quad\dot{T} = \frac{\hbar\kappa\gamma\sqrt{\Delta}}{4} |\nu| 
\end{equation}
and Friedmann equation
\begin{equation}
\dot{\nu}^2 = \frac{9}{\gamma^2\Delta}\left(A\nu^2+B-C^2 \frac{1}{\nu^2}\right)\,,
\end{equation}
where $A, B$ and $C$ are defined by
\begin{equation}
    A = \frac{\gamma^2\Delta}{3}\tilde{\Lambda}(\phi)\left(1-\frac{\gamma^2\Delta}{3}\tilde{\Lambda}(\phi)\right)\,,\quad B= C\left(1-\frac{2\gamma^2\Delta}{3}\tilde{\Lambda}(\phi)\right)\,,\quad C = \frac{8\pi_\phi^2}{3 \kappa \hbar^2}\,,
\end{equation}
were derived previously, but a key difference is that the momentum $\pi_\phi$ is no longer conserved so $A, B$ and $C$ all become functions of time. As we had previously noted in (\ref{standardeffFr}), by introducing the energy density
\begin{equation}
    \rho = \frac{\dot\phi^2}{2}+\frac{\tilde\Lambda}{\kappa} = \frac{8}{\kappa^2\hbar^2\gamma^2\,\Delta}\frac{\pi_\phi^2}{\nu^2}+\frac{\tilde\Lambda}{\kappa}\,,
\end{equation}
the effective Friedmann equation can be recast in a more familiar form as
\begin{equation}
    \left(\frac{\dot\nu}{3\nu}\right)^2 = \frac{\kappa}{3} \rho\left(1-\frac{\kappa\gamma^2\Delta}{3}\rho\right)\,.
\end{equation}
In this section all plots are obtained by integrating these equations numerically. For simplicity, we always start the integration at the bounce ($t=0$), where $\dot\nu=0$, or equivalently $\rho = \frac{3}{\kappa\gamma^2\Delta} \equiv \rho_c$. We will set the contribution from a cosmological integration constant $\Lambda$ to zero, given that dark energy plays no role in understanding the early inflationary period.

The remaining free initial conditions are $\nu(0)$ (which simply sets an arbitrary scale, and is set to one throughout), $\phi(0)$, the sign of $\pi_\phi(0)$ or of $\dot\phi$ (their magnitude can be obtained by setting $\rho(0)=\rho_c$) and $T(0)$, which we always set to $0$ with no loss of generality. 

The question of the origin of suitable initial conditions for inflation has been discussed in many papers and has attracted a certain amount of controversy (see, e.g., \cite{Ijjas:2013vea}). In standard approaches it is rather difficult to formulate the question rigorously, given that it is not clear when (and even less {\em how}) initial conditions are chosen, and that the space of initial conditions does not have a natural definition of measure with respect to which probabilistic statements could be made. One classic result suggests that the probability of finding initial conditions that yield sufficient amounts of inflation is exponentially suppressed \cite{Gibbons:2006pa}, which would render the usefulness of inflation as explaining the origin of our Universe questionable. Similar calculations in loop quantum cosmology have reached the opposite result \cite{Ashtekar:2009mm,Ashtekar:2011rm}, namely, that initial conditions leading to sufficient amounts of inflation are extremely likely. The setup here is different because initial conditions are set at the bounce (which is what we shall also do here) and the parameter space of initial data is compact, given the fixed (Planckian) value of the energy density at the bounce. The results are also obtained for the specific case of a quadratic potential. In general one has to be careful in explaining where exactly particular choices of parameter values and initial conditions are supposed to originate.

Standard calculations in cosmology for extracting a primordial power spectrum from a given inflationary model \cite{Baumann_2022} are based on assuming the validity of a slow-roll approximation and on choosing the Bunch--Davies vacuum as the quantum state for linearised scalar perturbations. This leads to predictions for observationally relevant parameters, in particular the amplitude $A_s$ and spectral tilt $n_s$ of scalar perturbations, which can be compared with observations such as those made by PLANCK \cite{Planck:2018jri}. It should be borne in mind that the extrapolation of observational data to a corresponding primordial power spectrum is model-dependent, so discussions in loop quantum cosmology would usually assume that the standard ($\Lambda$CDM) cosmology is unchanged. One question is then how the choice of initial conditions at the bounce is related to observationally viable parameter values during the following inflationary period. What is constrained by observations is typically the value of the inflaton field and the value of its potential energy at a time when an observationally relevant range of modes exit the Hubble horizon \cite{Baumann_2022}. Together, these constraints allow for constraints on free parameters of the model itself; in the simplest case where the inflationary potential has a single free parameter, it should be fixed by determining $n_s$ and $A_s$. If initial conditions are set at the bounce, as seems natural in a bouncing cosmology, converting these into the relevant field value at horizon crossing is an additional nontrivial step.

Traditionally, a lot of work in this area is based on choosing a quadratic potential
\begin{equation}
    V(\phi)=\frac{1}{2}m^2\phi^2\,.
\end{equation}
Observations of $n_s$ and $A_s$ would fix the mass $m$ and the value of $\phi$ at horizon crossing, and this is used to set a particular value for $m$ in \cite{Ashtekar:2009mm,Ashtekar:2011rm} as well as in many related works. PLANCK observations now strongly disfavour the quadratic potential, so if one is interested in an observationally viable cosmology alternative models are needed.

In loop quantum cosmology, kinetically dominated bounces are usually preferred, mostly because they can lead to imprints of quantum-gravitational corrections in the observable range of wavelengths \cite{Agullo:2013ai,Ashtekar:2021kfp}. From a theoretical point of view, a kinetically dominated bounce also allows approximating the inflaton by a free massless scalar field, which is the standard matter needed as a clock in order to define the theory. In unimodular loop quantum cosmology, one can find alternative scenarios in which a wider range of initial conditions can be considered: in general, the dynamics of inflation in a period before observationally accessible modes are generated are not strongly constrained by observations. There is substantial freedom in choosing features of the inflationary potential away from the eventual slow-roll regime that is observationally required. In a quantum-gravity inspired bounce scenario, there is presumably a wide range of options for choosing an inflationary potential and initial conditions at the bounce while both being consistent with CMB observations and generating corrections to standard inflation that might become observationally relevant (at low multipoles or long wavelengths).    

\

More precisely, the requirement that perturbations are stretched enough to reproduce the observed nearly scale-invariant primordial power spectrum sets a lower bound on the number of $e$-folds of inflation (around 60 assuming the post-inflationary history of standard cosmology), while the demand that there are possible imprints of quantum gravity effects in the observable window of the power spectrum sets an upper bound on the number of $e$-folds: the Universe cannot inflate too much, or any features of the power spectrum will be stretched out of the observable window into the infrared. 

Furthermore, without having to actually compute primordial power spectra, with the background evolution alone it is possible to find indications of whether particular cases may leave imprints within the observable window. Classically, the conformal time evolution of Fourier modes of the gauge-invariant perturbation $u_k(\eta)$ is governed by the equation of motion of a harmonic oscillator with time-dependent mass $s(\eta)$:
\begin{equation}
    u^{\prime\prime}_k(\eta) + \left[k^2+s(\eta)\right] u_k(\eta) = 0\,.
\end{equation}
For scalar (Mukhanov--Sasaki) modes $s(\eta) = -z^{\prime\prime}/z$ with $z = a \phi'/H_c$, where $H_c=a'/a$ is the conformal Hubble parameter\footnote{Related to the Hubble parameter $H=\dot{a}/a$, where one differentiates with respect to cosmic time, as $H_c=aH$.}. During slow roll, this can be approximated by $s(\eta) \approx -a^{\prime\prime}/a$, which in a quasi-de Sitter phase can be further written as $s(\eta) \approx - 2 H_c^2$. In this regime, one obtains the usual interpretation that modes inside the comoving Hubble horizon $R_H = 1/H_c$ (with $k \gg H_c$) show an oscillatory behaviour, whereas modes that have crossed the horizon ($k \ll H_c$) ``freeze out'', tending to a constant value. The usefulness of inspecting the quantity $H_c$ is therefore evident: it allows us not only to determine when the Universe undergoes epochs of accelerated (increasing $H_c$) or decelerated (decreasing $H_c$) expansion, but also to determine when the observable modes oscillate and freeze out. In the standard picture, modes are fixed to the Bunch--Davies vacuum at the onset of inflation and then evolve by oscillating until they gradually cross out during inflation and remain frozen until well after inflation ends.

A deviation from standard results can therefore arise when pre-inflationary dynamics affect observable modes (for example if they cross out and in before standard inflation begins), such that it is no longer reasonable to assume they reach the onset of inflation in the Bunch--Davies vacuum. However, one should keep in mind that $H_c$ is only the relevant quantity during a quasi-de Sitter expansion, and that more generally the ``horizon'' to compare $k$ with is rather $\sqrt{|s(\eta)|}$. This is particularly important when considering pre-inflationary dynamics and possible impacts of background dynamics on observable modes.

For a given quantum theory, one may then find corrections to the time-dependent mass $s(\eta)$ that capture the main effects of the quantum dynamics. These have been obtained for loop quantum cosmology in the hybrid and dressed metric approaches \cite{ElizagaNavascues:2017avq}. In earlier works, where the actual form of $s(\eta)$ was not yet available, authors often compared the curvature radius with physical wavelengths (or equivalently, the inverse of the comoving curvature radius with comoving wavenumbers $k$), following the intuition that modes are affected differently depending on the magnitude of their physical wavelengths with respect to the curvature radius of the background. As we will see, this approach does not always match considering $\sqrt{|s(\eta)|}$ of different quantum formalisms.

For the remainder of this section, we again choose units in which $\hbar=G=1$ and hence $\kappa=8\pi$, so that all energy scales are given in units of the Planck mass $m_{{\rm Pl}}=\sqrt{\hbar/G}$. This is in agreement with literature in loop quantum cosmology such as \cite{Ashtekar:2009mm,Ashtekar:2011rm} but differs from a lot of cosmology literature in which the reduced Planck mass $M_{{\rm Pl}}=\sqrt{\hbar/\kappa}$ is used, and one rather chooses $\kappa=1$. Again, we also choose $\gamma=0.2375$.

\subsection{Quadratic potential}

\begin{figure}
    \centering
    \includegraphics[width=0.5\linewidth]{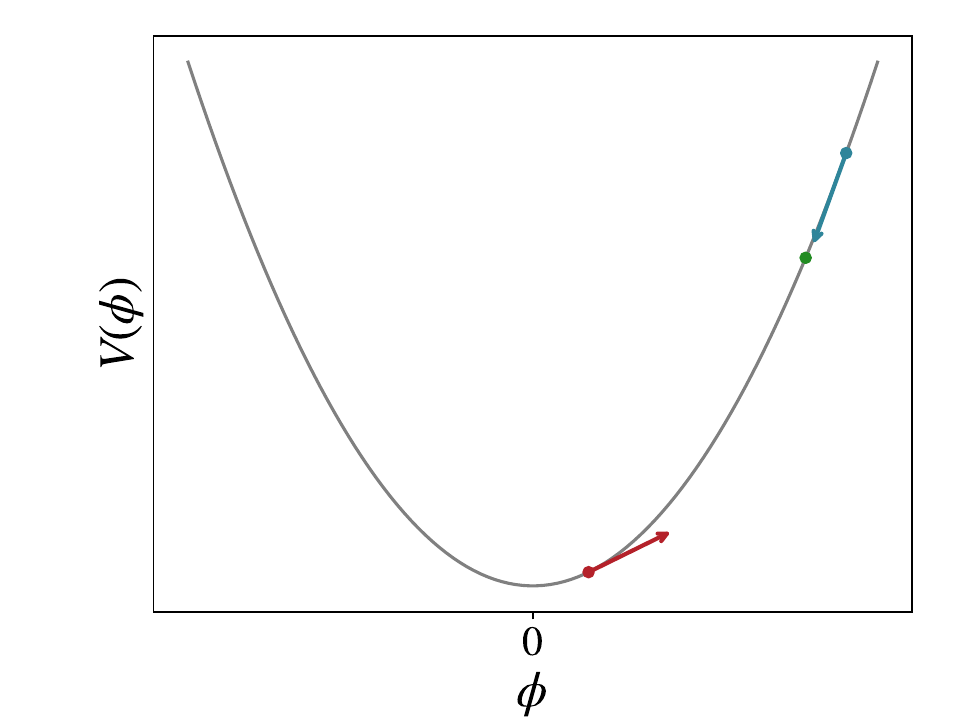}
    \caption{Illustrative initial conditions for a quadratic potential. Red and blue define kinetically dominated bounces and green defines a bounce dominated by potential energy. For kinetic dominance the sign of $\dot\phi$ is relevant and both signs are explored here, as represented by the arrows.}
    \label{fig:Vphi-quad}
\end{figure}

\begin{figure}[htbp]
    \centering
    \begin{subfigure}{0.45\textwidth}
        \centering
        \includegraphics[width=\textwidth]{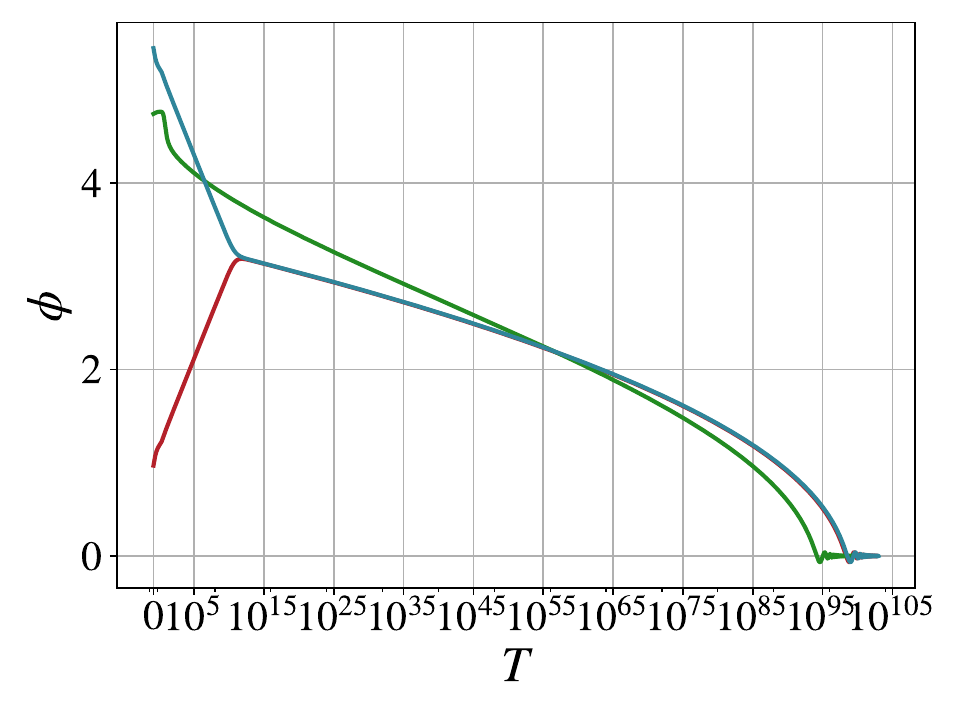}
    \end{subfigure}%
    \begin{subfigure}{0.45\textwidth}
        \centering
        \includegraphics[width=\textwidth]{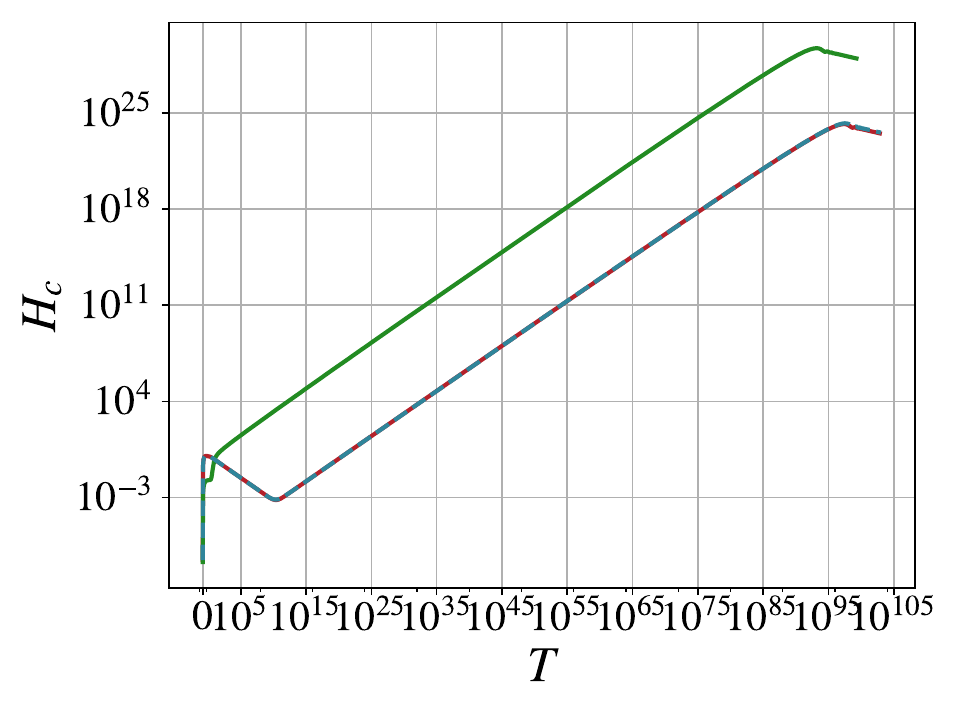}
    \end{subfigure}    
    \caption{Background dynamics for the quadratic potential as a function of unimodular time $T$ for the three cases colour-coded in Figure \ref{fig:Vphi-quad}. \textbf{Left:} scalar field $\phi(T)$. \textbf{Right:} background expansion rate $H_c=aH$ (red and blue overlap). We have chosen values of $m$ and $\phi_0$  to produce about 70 $e$-folds from the bounce to the end of inflation: \textbf{red:} $m = 1.2 \times 10^{-6}$, $\phi(0) = 0.97$; \textbf{blue:} $m = 1.2 \times 10^{-6}$, $\phi(0) = 5.45$; \textbf{green:} $m = 3.6 \times 10^{-2}$, $\phi(0) = 4.75$.}
    \label{fig:background-quad}
\end{figure}
This potential was often used in earlier studies of cosmological perturbations in loop quantum cosmology. There are three qualitatively different types of initial conditions:

\begin{itemize}
\item Kinetically dominated bounce with $\pi_\phi(0)>0$: represented in red in Figures \ref{fig:Vphi-quad} and \ref{fig:background-quad}. As illustrated in the plots of Figure \ref{fig:background-quad}, choosing initial conditions such that the kinetic energy dominates over the potential and pushing $\phi$ up the potential at the bounce leads to a very short period of very fast acceleration immediately after the bounce, followed by a period of decelerated expansion where the kinetic energy decreases dramatically, until the potential dominates and slow-roll inflation begins. Inflation ends once the inflaton reaches the minimum of the potential and begins oscillating. This is a typical case that has been studied in the literature; it has a range of observable modes that could in principle be sensitive to quantum corrections to the background evolution and to the differences between different approaches for including perturbations.
\item Kinetically dominated bounce with $\pi_\phi(0) <0$: represented in blue in Figures \ref{fig:Vphi-quad} and \ref{fig:background-quad}. As we can see from the background evolution, even though the field starts by being pushed down the potential, it quickly catches up to the previous case and reproduces essentially the same background dynamics. This case is equivalent to choosing $\phi(0) < 0$ and keeping $\pi_\phi (0) > 0$, as is usually done in the literature.
\item Potential-dominated bounce: represented in green in Figures \ref{fig:Vphi-quad} and \ref{fig:background-quad}. In this case the sign of $\pi_\phi(0)$ or $\dot\phi(0)$ is essentially irrelevant, and inflation starts almost immediately. For the bounce to be potential dominated, if $m^2$ is small in Planck units (as is required to fit observations even approximately), $\phi(0)$ needs to take a value far above the Planck scale at the bounce (recall that the total energy density at the bounce is fixed to be the Planckian energy density $\rho_c$). Such a large field value leads to more $e$-folds by the end of inflation, and such scenarios necessarily have a lower bound for the number of $e$-folds they generate. This observation is at the heart of the results of \cite{Ashtekar:2009mm,Ashtekar:2011rm} that, for a choice of $m$ such that agreement with observations is possible, all such initial conditions lead to enough $e$-folds of inflation. But this is ``too much'' inflation in the sense that the effects of pre-inflationary dynamics then fall very far infrared of the observational window.
\end{itemize}

Since the quadratic model is disfavoured by observations, it is perhaps not particularly useful for investigating the phenomenology of loop quantum cosmology.

\subsection{Starobinsky potential}

\begin{figure}
    \centering
    \includegraphics[width=0.5\linewidth]{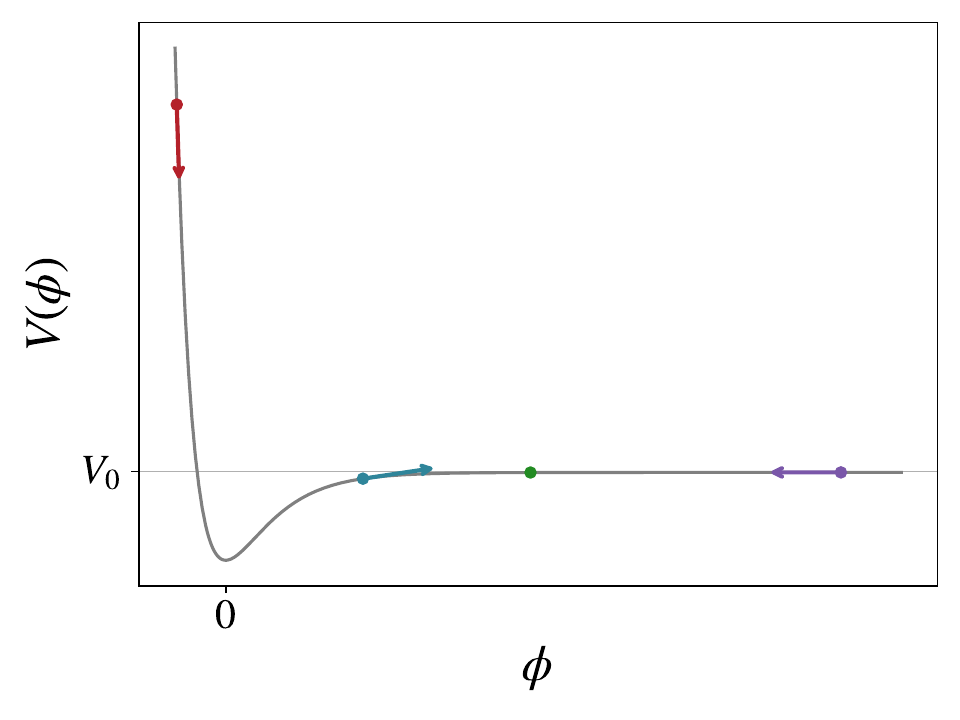}
    \caption{Illustrative initial conditions for the Starobinsky potential. Red, blue and purple define kinetically dominated bounces and green defines a potential-dominated one. For kinetic dominance and $\phi(0)>0$ the sign of $\dot\phi$ is relevant and both options are represented by the arrows.}
    \label{fig:Vphi-star}
\end{figure}

A more phenomenologically relevant case is given by the Starobinsky potential
\begin{equation}
    V(\phi) = V_0 \left(1-e^{-\sqrt{\frac{2\kappa}{3}}\phi}\right)^2\,.
\end{equation}
We have again considered different types of initial conditions as illustrated in Figure \ref{fig:Vphi-star}.

In this case the potential is no longer even in $\phi$, and positive and negative field values behave very differently. In particular, we do not have a scenario with potential domination and initially negative field value. That is because, due to the limit $\rho \leq \rho_c$, and the steepness of the potential for $\phi<0$, there is no choice of initial conditions that leads to anywhere near enough $e$-folds of inflation. For a fixed $V_0$, more inflation is obtained for larger $|\phi(0)|$, but since the potential is very steep on the left, the limit of $V(\phi) = \rho_c$ is reached for relatively small $|\phi|$, and no more inflation can be obtained. One has to decrease $V_0$ to be able to reach larger $|\phi|$, but cannot decrease it enough to obtain 50-60 $e$-folds before transitioning to a kinetically dominated bounce.

\begin{figure}[htbp]
    \centering
    \begin{subfigure}{0.45\textwidth}
        \centering
        \includegraphics[width=\textwidth]{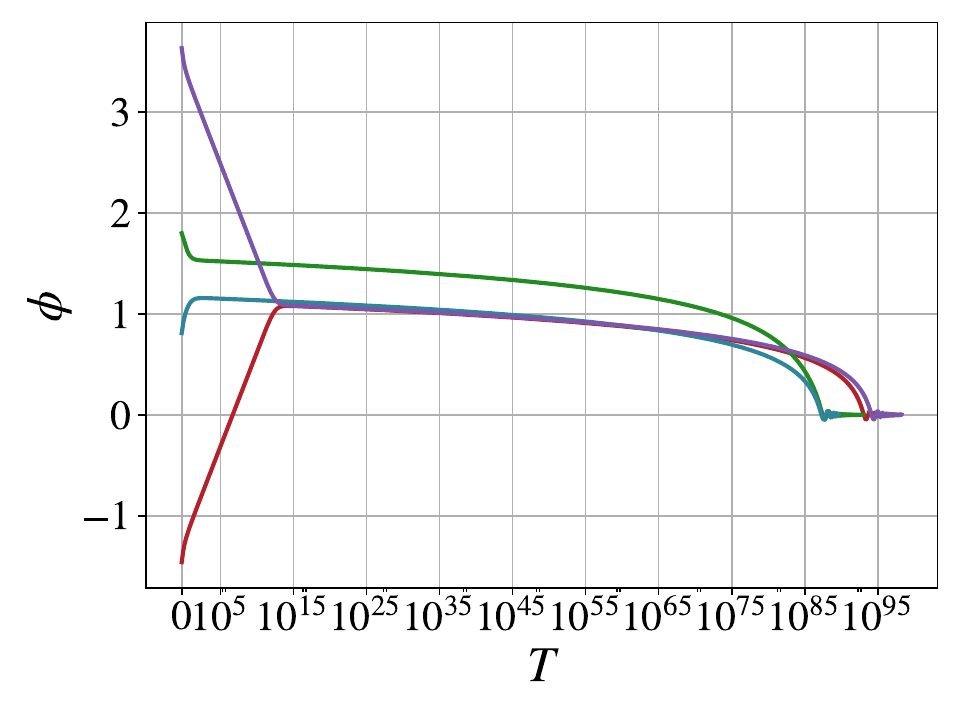}
    \end{subfigure}%
    \begin{subfigure}{0.45\textwidth}
        \centering
        \includegraphics[width=\textwidth]{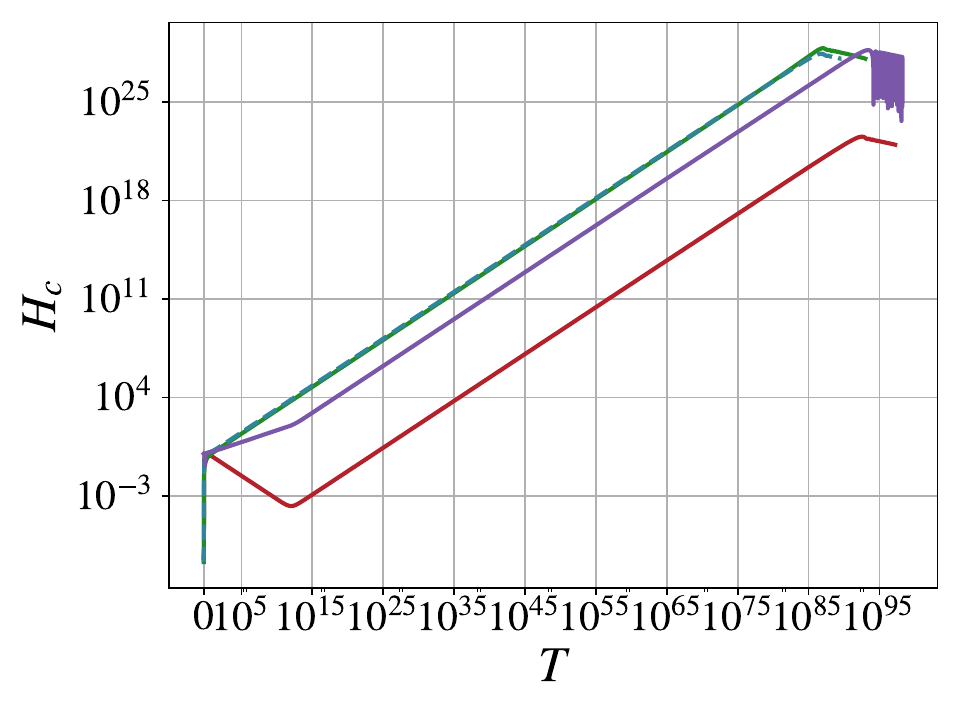}
    \end{subfigure}    
    \caption{Background dynamics for the Starobinsky potential as a function of unimodular time $T$ for the five cases colour-coded in Figure \ref{fig:Vphi-star}. \textbf{Left:}  scalar field $\phi(T)$. \textbf{Right:}  background expansion rate $H_c=aH$.  $V_0$ and $\phi(0)$ are chosen to produce about 66 $e$-folds from the bounce to the end of inflation: \textbf{red:} $V_0 = 1.77 \times 10^{-13}$, $\phi(0) = -1.46$; \textbf{green:} $V_0 = 0.35$, $\phi(0) = 1.80$; \textbf{blue:} $V_0 = 0.1$, $\phi(0) = 0.81$; \textbf{purple:} $V_0 = 1.77 \times 10^{-13}$, $\phi(0) = 3.634$.}
    \label{fig:background-star}
\end{figure}

Similarly, there is no representation in Figure \ref{fig:Vphi-star} for a kinetically dominated bounce with $\phi(0) < 0$ and climbing up the potential. Given the previous explanation it is easy to understand why: such a scenario would evolve by pushing the field up the potential until it reaches a maximum and turns around, at which point this is essentially equivalent to dropping it from that point as in a potential dominated bounce. The remaining dynamical phase is not long enough to generate enough $e$-folds, which would need to be obtained in the first part of the dynamics before the field turns around. Since the potential is very steep and the total energy density is again bounded by the critical density, it is also impossible to achieve enough $e$-folds in this scenario.

Figure \ref{fig:Vphi-star} represents the types of initial conditions that may lead to enough inflation:

\begin{itemize}

\item Kinetically dominated bounce with $\phi(0) < 0$: represented in red in Figures \ref{fig:Vphi-star} and \ref{fig:background-star}. Figure \ref{fig:background-star} shows that in this case the inflaton starts by travelling down the potential with enough velocity that it dominates over the potential energy, then crosses over the minimum reaching the plateau on the positive side, until its velocity eventually decreases enough that the potential energy takes over and it starts slowly rolling down the potential towards the minimum, generating the standard inflationary phase. This leads to a very fast period of super-inflation right after the bounce, followed by a period of decelerated expansion, all before the potential takes over and standard inflation begins. These different dynamics of the background are appealing as they may easily leave imprints in some region of the observable window. 

\item Potential-dominated bounce with $\phi(0) > 0$: depicted in green in Figures \ref{fig:Vphi-star} and \ref{fig:background-star}. As shown in Figure \ref{fig:background-star}, with this choice of initial conditions, slow-roll inflation starts right after a very short period of super-inflation after the bounce. For such initial conditions, one has the opposite problem as for negative $\phi(0)$: for a value of $V_0$ that is compatible with observations ($V_0 \sim 10^{-13}$), and given the shallowness of the potential on the right, for the bounce to be potentially dominated $\phi(0)$ needs to be so big that we generate too many $e$-folds of inflation and any pre-inflationary effects will not fall in the observable window. To limit the number of $e$-folds one would need to choose a much larger (Planck-scale) $V_0$, as we have done for the case represented in the figures, which then is not compatible with observations of $A_s$ and $n_s$.

\item Kinetically dominated bounce with $\phi(0) > 0$ and $\pi_\phi(0) > 0$: represented in blue in Figures \ref{fig:Vphi-star} and \ref{fig:background-star}. In this case, the kinetic energy has to be large enough for $\rho \approx \dot\phi^2/2 = \rho_c$ at the bounce. So, $\dot\phi(0)$ is necessarily large, which for $\phi(0)>0$ pushes the field very far up the potential and, since the slope is so low on this side, this creates a very long period of slow roll and generates too much inflation. The lower we make $\phi(0)$ (to compensate for this), the higher we need to make $\dot\phi(0)$ to still saturate $\rho=\rho_c$ at the bounce, which then does not succeed in generating fewer $e$-folds. The only solution is again to make $V_0$ much larger. Unfortunately this means this case is also not observationally relevant, but it is nonetheless interesting that it leads to very similar background dynamics to the previous case.

\item Kinetically dominated bounce with $\phi(0) > 0$ and $\pi_\phi(0) < 0$: represented in purple in Figures \ref{fig:Vphi-star} and \ref{fig:background-star}. Since now the field is pushed down the potential, the problem of the previous case does not arise and we can easily get just enough $e$-folds of inflation such that effects of the quantum dynamics may still lie in the observable window. However, this case produces less dramatic changes to the horizon with respect to standard inflation than the first (red) case, which suggests that the quantum-corrected dynamics leave weaker  observable imprints.

\end{itemize}

It is important to point out that the conclusions of \cite{Ashtekar:2009mm,Ashtekar:2011rm} for a quadratic potential, namely that a potential-dominated bounce always leads to enough inflation whereas a kinetically dominated one only does so for a portion of parameter space, do not translate directly to other potentials. For the Starobinsky model, we find that the potential-dominated case with $\phi(0)<0$ never leads to enough inflation, and the kinetically dominated one is similarly restricted for a field initially climbing up the potential. Conversely, on the $\phi(0)>0$ side it remains the case that a potential-dominated bounce always leads to enough inflation if $V_0$ is fixed observationally, whereas a kinetically dominated one only does so for a certain region of parameter space. These considerations would presumably affect the probability for obtaining sufficient inflation overall. If a different and more complicated potential is chosen, such conclusions would again have to be revisited. This is not surprising as in general the question of whether certain initial conditions lead to ideal background evolutions (enough $e$-folds, but not too many) strongly depends on the form of the potential.

\subsection{$\alpha$-attractor potential}

\begin{figure}
    \centering
    \includegraphics[width=0.5\linewidth]{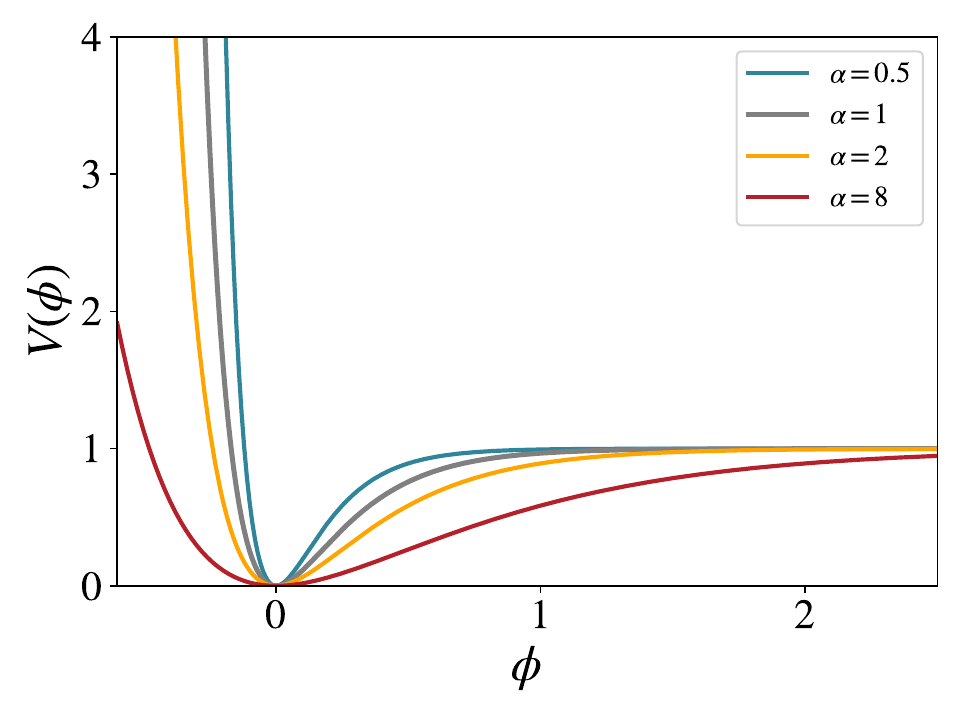}
    \caption{$\alpha$-attractor potential for different choices of $\alpha$. $\alpha=1$ is the Starobinsky potential.}
    \label{fig:Vphi-different-alphas}
\end{figure}

The PLANCK results are compatible with a large number of inflationary potentials. A particularly interesting class of models, which can be seen as an extension of the Starobinsky potential, are known as $\alpha$-attractors \cite{Kallosh:2013yoa,Kallosh:2015lwa}. Here we focus on the ``E-model'' given by
\begin{equation}
    V(\phi) = V_0 \left(1-e^{-\sqrt{\frac{2\kappa}{3\alpha}}\phi}\right)^2
\end{equation}
where $\alpha$ is a parameter which can deviate significantly from the Starobinsky value $\alpha=1$. Figure \ref{fig:Vphi-different-alphas} illustrates the shape of potentials for different values of $\alpha$; we will use $\alpha = 8$. As we will see, this is a model that is compatible with our observations of the power spectrum while also allowing for a wider range of phenomenologically interesting initial conditions within loop quantum cosmology. The main reason for this is that the predictions of such models for observable parameters $A_s$ and $n_s$, and with a tensor-to-scalar ratio small enough to evade current upper bounds, are only weakly dependent on the parameter $\alpha$ (the PLANCK constraint \cite{Planck:2018jri} is $\alpha<10^{1.3}\approx 20$) while the potential can deviate strongly from the Starobinsky model at higher energy scales that are unconstrained by observation, where corrections such as those in loop quantum cosmology become important.

\begin{figure}
    \centering
    \includegraphics[width=0.5\linewidth]{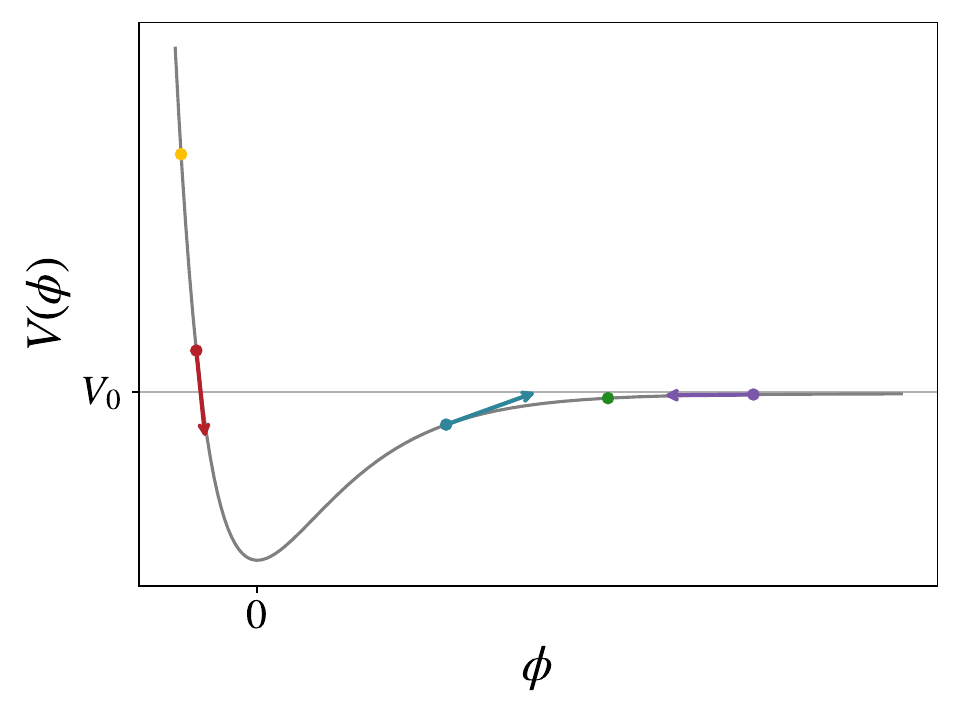}
    \caption{Illustrative initial conditions for the $\alpha$-attractor potential with $\alpha = 8$. Red, blue and purple define kinetically dominated bounces, yellow and green define a potential-dominated one. For kinetic dominance the sign of $\dot\phi$ is represented by the arrows. The particular values of $V_0$ and $\phi(0)$ are chosen within those constraints to produce about 68 $e$-folds from the bounce to the end of inflation: \textbf{yellow:} $V_0=1.17 \times 10^{-12}$, $\phi(0) = -9.00$; \textbf{red:} $V_0 = 1.17 \times 10^{-12}$, $\phi(0) = -0.58$; \textbf{green:} $V_0 = 0.35$, $\phi(0) = 2.99$; \textbf{blue:} $V_0 = 0.1$, $\phi(0) = 1.61$; \textbf{purple:} $V_0 = 1.17 \times 10^{-12}$, $\phi(0) = 4.23$. (See, e.g., \cite{Kallosh:2025ijd} on observational constraints on $\alpha$-attractors.)}
    \label{fig:Vphi-alpha}
\end{figure}

\begin{figure}[htp]
    \centering
    \begin{subfigure}{0.45\textwidth}
        \centering
        \includegraphics[width=\textwidth]{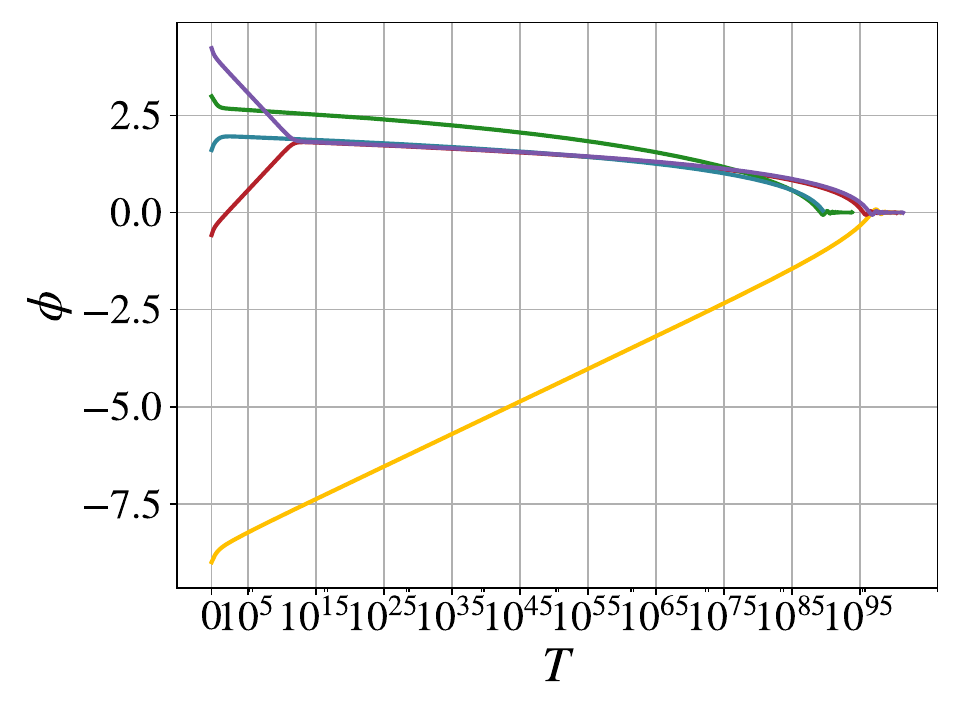}
    \end{subfigure}%
    \begin{subfigure}{0.45\textwidth}
        \centering
        \includegraphics[width=\textwidth]{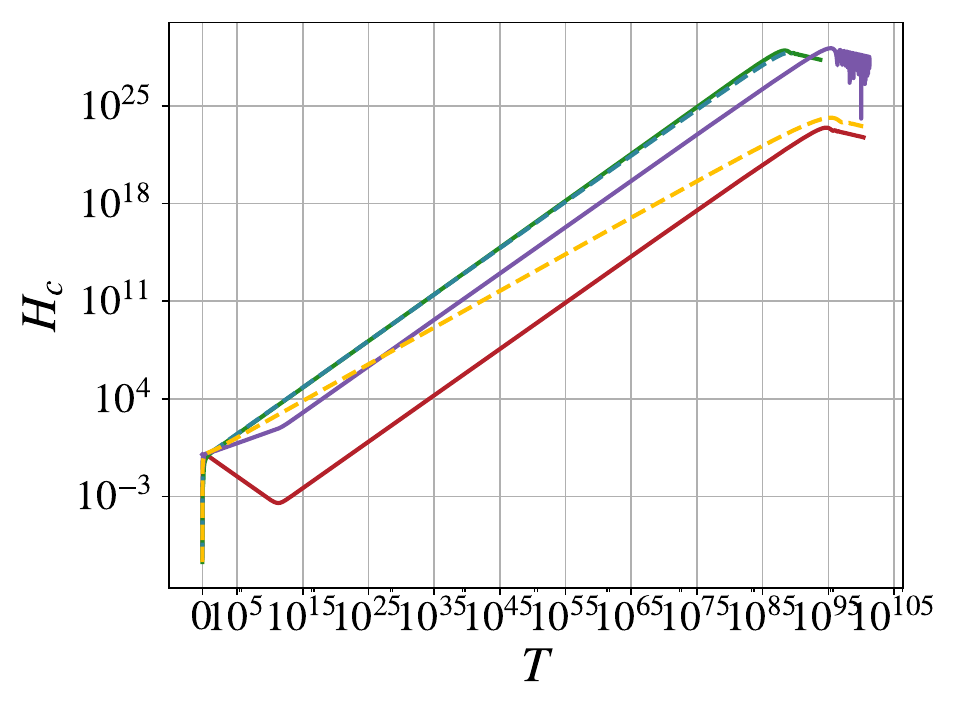}
    \end{subfigure}    
    \caption{Background dynamics for the $\alpha$-attractor as a function of unimodular time $T$ for the five cases colour-coded in Figure \ref{fig:Vphi-alpha}. \textbf{Left:} scalar field $\phi(T)$. \textbf{Right:} expansion rate $H_c=aH$.}
    \label{fig:background-alpha}
\end{figure}

The $\alpha$-attractor potential allows for the types of initial conditions we already encountered for the Starobinsky potential, represented in the same red, green, blue and purple in Figures \ref{fig:Vphi-alpha} and \ref{fig:background-alpha}. However, due to the different steepness particularly for negative $\phi$, we are now also able to set observationally relevant initial conditions with potential dominance at the bounce, as represented in yellow. Because the potential is less steep, even with a low value of $V_0$ (chosen to be the same as for the red and purple scenarios), we can increase $|\phi(0)|$ enough to achieve potential dominance while remaining within $\rho\le\rho_c$. This allows for the desired number of $e$-folds (sufficiently many but not too many) from the bounce to the end of inflation. There is a short period of super-inflation, but it transitions seamlessly to slow-roll inflation without going through a decelerated phase.

The $\alpha$-attractor with such initial conditions provides an example of a realistic inflationary scenario (consistent with PLANCK observations) in which the scalar field potential  is always important, even at the bounce. This scenario produces just enough inflation so that imprints from the quantum dynamics may fall within the observational window, but since the bounce is dominated by potential energy, the usual procedure of ignoring the potential close to the bounce to proceed with quantisation with $\phi$ as time is not applicable. Models based on unimodular time provide an alternative approach where such a scenario can be studied consistently.

\begin{figure}[htp]
    \centering
    \begin{subfigure}{0.47\textwidth}
        \centering
        \includegraphics[width=\textwidth]{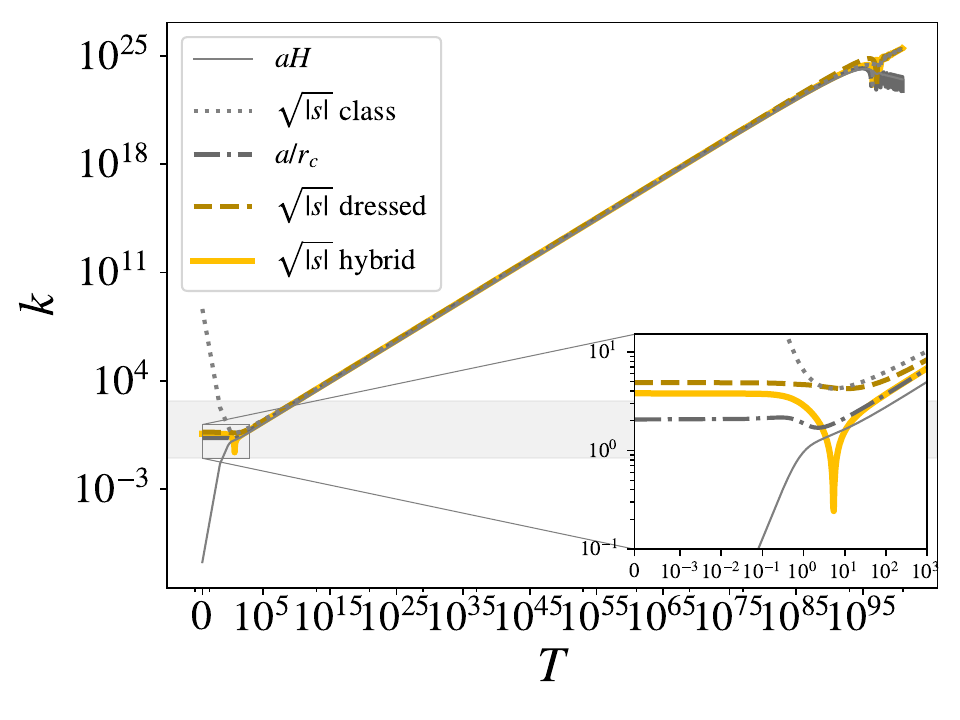}
    \end{subfigure}%
    \begin{subfigure}{0.47\textwidth}
        \centering
        \includegraphics[width=\textwidth]{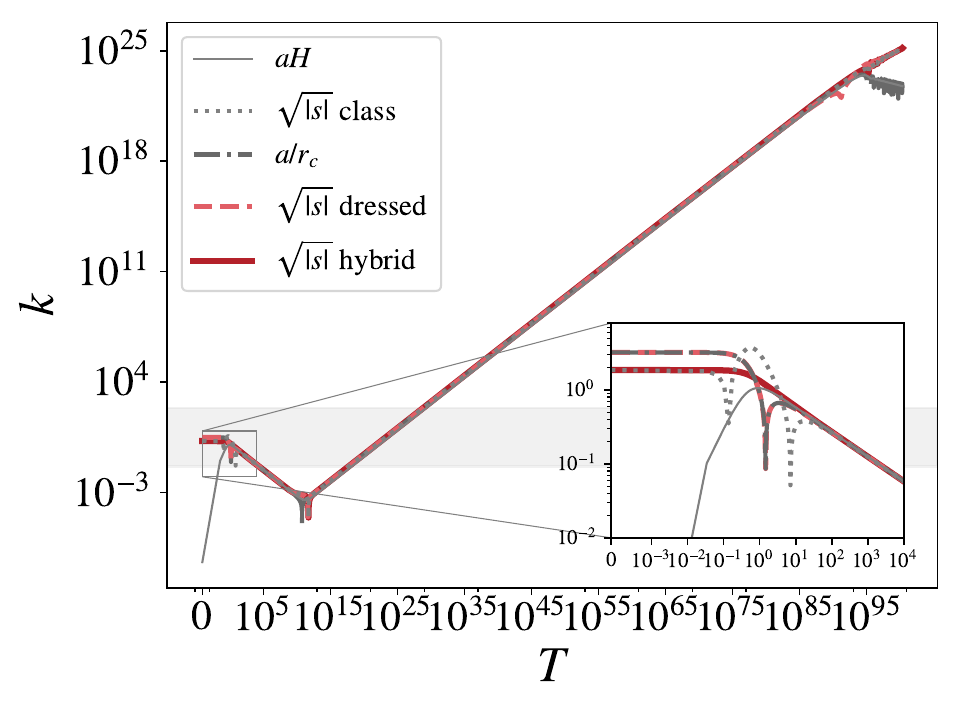}
    \end{subfigure}
    \begin{subfigure}{0.47\textwidth}
        \centering
        \includegraphics[width=\textwidth]{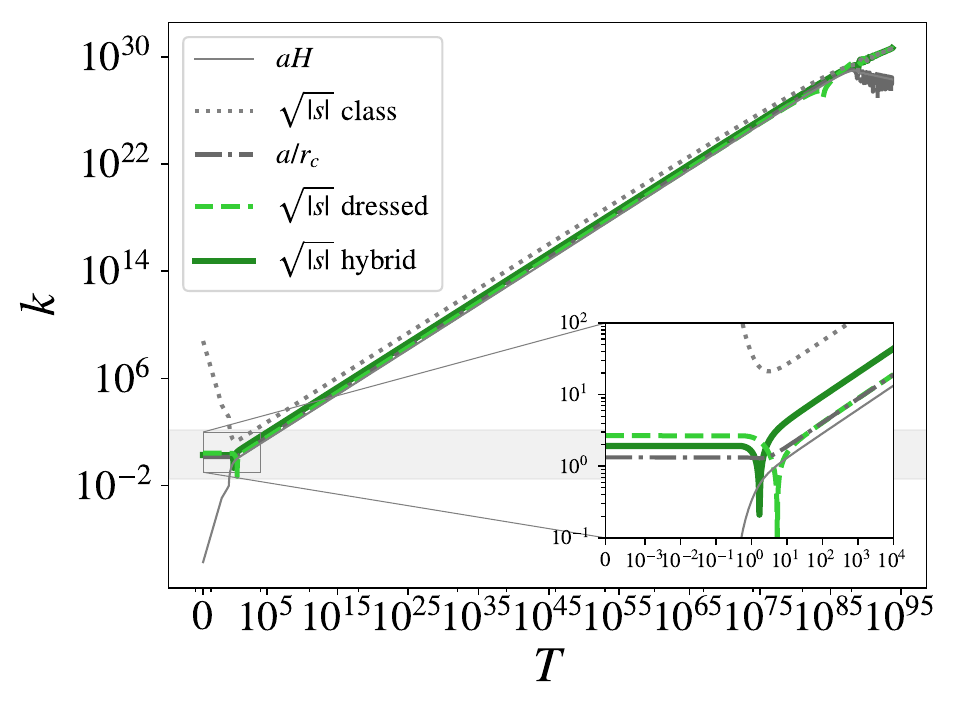}
    \end{subfigure}%
    \begin{subfigure}{0.47\textwidth}
        \centering
        \includegraphics[width=\textwidth]{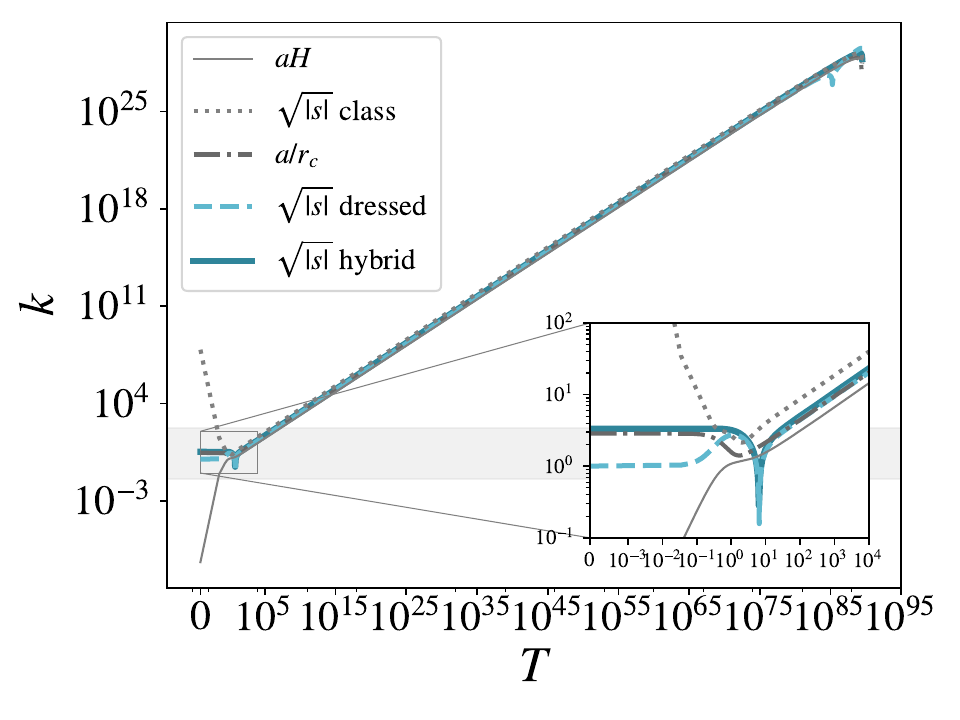}
    \end{subfigure}
    \begin{subfigure}{0.47\textwidth}
        \centering
        \includegraphics[width=\textwidth]{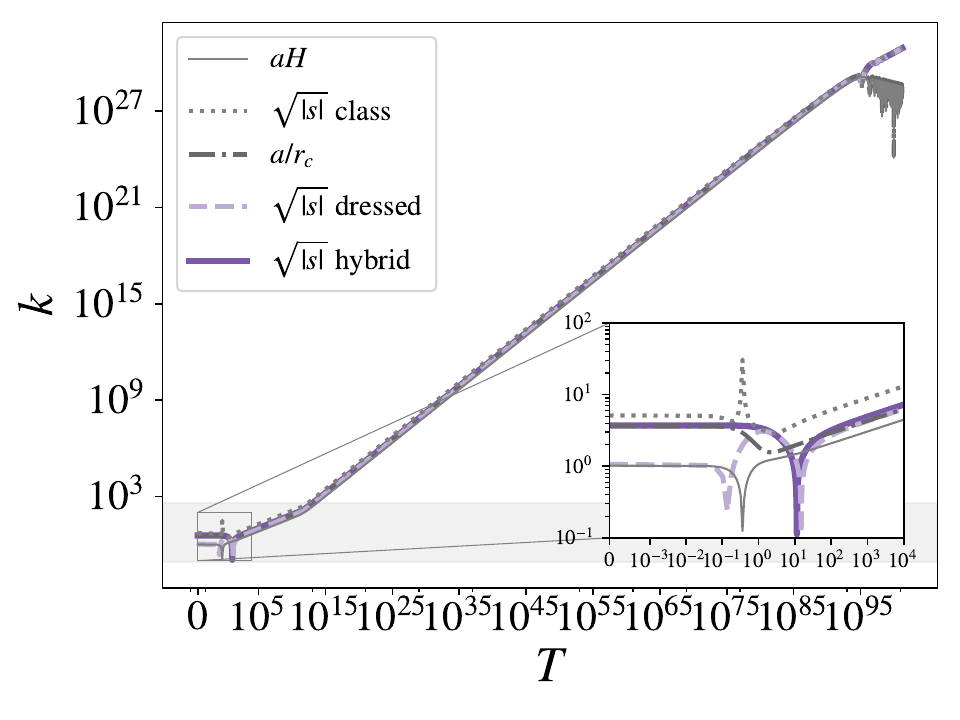}
    \end{subfigure} 
    \caption{Evolution of the time-dependent mass for Mukhanov--Sasaki modes of different formalisms, as well as classical quantities for comparison, for the five cases colour-coded in Figure \ref{fig:Vphi-alpha}. The observable range of modes for each case is shown as the shaded grey region.}
    \label{fig:s-alpha}
\end{figure}

In Figure \ref{fig:s-alpha}, we illustrate how one can infer possible effects of quantum corrections on the primordial power spectrum purely from knowing the background dynamics. In all five plots, the grey region represents the band of observable comoving modes.\footnote{The observable window given by PLANCK corresponds to the physical wavenumber range today of $\left[10^{-4},0.5 \right]\,\textrm{Mpc}^{-1}$. This range is translated to physical wavenumbers at the bounce (which corresponds to our comoving wavenumbers $k$, as we normalise the scale factor to be $1$ at the bounce) by multiplying it by $e^{N_T}$ where $N_T$ is the total number of $e$-folds the Universe has undergone from the bounce until today. Our numerical calculations provide us with that number up until the end of inflation. To account for the rest of the evolution until today we assume the Universe has expanded a further $70$ $e$-folds, a typical estimate to account for the standard post-inflationary history. Finally, the region shown in grey is obtained by converting $\textrm{Mpc}^{-1}$ into Planck units.} Because we choose initial conditions that always lead to about 68 $e$-folds of inflation, this band always encompasses the modes that are comparable to the size of the horizon early on. In the plots we compare the actual horizon scale (derived from $\sqrt{|s|}$, see above) of the hybrid and dressed metric approaches with its classical counterpart, with the inverse of the comoving Hubble horizon (which matches the classical counterpart when in de Sitter), and with the inverse of the comoving curvature radius. The expressions for $\sqrt{|s|}$ used for the hybrid and dressed metric are those derived in the literature for standard loop quantum cosmology. It would be important to check whether similar expressions can be obtained in a unimodular framework. Figure \ref{fig:s-alpha} should be seen as a qualitative analysis for when known quantum corrections in loop quantum cosmology might become relevant; a full analysis in unimodular loop quantum cosmology, which we must leave for future work, would allow connecting to studies of cosmological perturbations in loop quantum cosmology such as \cite{Zhu:2017jew,Li:2018vzr,Agullo:2023rqq}.

Note that the curvature radius is not always in agreement with the horizon of either approach, unlike what was suggested in early literature. In all five scenarios the predictions differ between the hybrid and dressed metric approaches, so these could in principle affect observable modes differently and leave specific imprints in the primordial power spectrum. Specifically, this is true in the observationally relevant scenarios shown in yellow, red and purple. Note also that the qualitative horizon evolution is different in these three scenarios, and could affect a range of observable modes differently: in the yellow case (potential-dominated bounce), the horizon increases after the initial period of super-inflation, meaning that modes that have crossed the horizon will remain frozen until the end of inflation, whereas in the red case the horizon still decreases during the period of kinetic dominance, meaning some modes might cross back in and out before remaining frozen for the rest of inflation. This is a concrete example of an observationally relevant model that could not be quantised in loop quantum cosmology with the standard choice of $\phi$ as clock (constructions in full loop quantum cosmology beyond effective equations require neglecting the potential $V(\phi)$ near the bounce, as in \cite{Agullo:2013ai}), and that is potentially observationally distinct from other configurations.

\section{Conclusions}
\label{conclsec}

We have discussed different motivations for studying loop quantum cosmology within unimodular gravity, where an additional global degree of freedom conjugate to the cosmological constant can be used to define a standard of time. Given that the definition of unimodular time depends on a choice of foliation, such a proposal is particularly natural in situations where a preferred foliation exists, such as homogeneous cosmology. We followed the pioneering work of \cite{Chiou:2010ne} in introducing a prescription for loop quantum cosmology in unimodular gravity within the usual symmetry-reduced setting. We remained at the semiclassical level of effective equations, in agreement with much of the literature on loop quantum cosmology. More systematic studies of the quantum theory could show how effective equations arise from particular semiclassical states, and also address the important issue of inequivalence of quantum theories defined with respect to different clocks \cite{Gielen:2020abd}, which cannot be seen at the level of effective equations.

We were able to derive various analytical solutions for different cases, some of which are new, but these were mostly restricted to the case of a constant scalar field potential (effectively another constant contribution to the cosmological constant). We then moved into a more phenomenological direction, investigating different choices of inflationary potential and possible scenarios for initial conditions at the bounce. While this part of the work is similar to other papers that do not use unimodular time, we would stress the usefulness of unimodular time in no longer requiring an approximately free massless scalar field to be used as a clock. In particular, we showed a possible realistic scenario for an $\alpha$-attractor potential, where the bounce is dominated by potential energy in the scalar field. Such a scenario could be studied in a quantum theory based on a unimodular choice of clock but not in the standard approach. We hope to initiate further studies into initial conditions in inflation within loop quantum cosmology; the evidence from the three possible cases we considered (quadratic, Starobinsky, $\alpha$-attractor) suggests that what types of initial conditions are possible or phenomenologically appealing, or which ones might be considered natural or likely, is highly model-dependent. This observation does not seem surprising, given that these questions probe the details of inflation at the Planck scale, which is not strongly constrained by existing observations. As usual, one may consider this model dependence to be either a feature or a significant problem for inflation. Within loop quantum cosmology, where the choice of inflationary potential is not constrained by fundamental quantum gravity considerations, it seems to open the door to a wider range of phenomenological models than have been considered so far, with potentially distinct observational signatures.

An open question in loop quantum cosmology is how it relates to the full theory of loop quantum gravity, which is still under development. Outside of homogeneous cosmology, while the introduction of unimodular time means there is always a standard of time that can be used to define evolution, it remains to be seen how much a unimodular approach helps in completing the full theory to a point where an extraction of cosmological phenomenology is possible. A more modest goal might be to develop other symmetry-reduced models for loop quantum gravity, such as anisotropic Bianchi or spherically symmetric black hole models, within a unimodular approach. Such constructions could again be used to illuminate technical, conceptual and possible phenomenological aspects of unimodular gravity within loop quantum gravity.

{\em Acknowledgements.} --- The research leading to these results was  funded by the Royal Society through the University Research Fellowship Renewal URF$\backslash$R$\backslash$221005.

\bibliographystyle{JHEP}
\bibliography{bib}

\end{document}